\documentclass[times,twocolumn,final,longtitle]{elsarticle}
\usepackage{jasr}
\usepackage{framed,multirow}
\usepackage{amssymb}
\usepackage{natbib}
\usepackage{latexsym}
\usepackage[normalem]{ulem}
\usepackage{url}
\usepackage{amsmath}
\usepackage{subcaption}
\usepackage{xcolor}
\definecolor{newcolor}{rgb}{.8,.349,.1}
\usepackage[citebordercolor=white]{hyperref}

\usepackage{graphicx}   % Including figure files
\usepackage{amsmath}    % Advanced maths commands
\usepackage{multicol} 

\journal{Advances in Space Research}

\begin{document}

%\linenumbers

%\verso{Given-name Surname \textit{etal}}
\verso{D. Debnath and H.-K. Chang}

\begin{frontmatter}

\title{Accretion Flow Properties of MAXI J1834-021 during Its Multi-peaked Outburst in 2023}
%\subtitle{Flow Properties of MAXI J1834-021}

%\author{Dipak Debnath\inst{1,2} \and Hsiang-Kuang Chang\inst{1,3} }

%\author[1]{Dipak \snm{Debnath}}
\author[1,2]{Dipak \snm{Debnath}\corref{cor1}}
\cortext[cor1]{Corresponding author: 
  email: dipakcsp@gmail.com}
\author[1,3,4,5]{Hsiang-Kuang \snm{Chang}}

\address[1]{Institute of Astronomy, National Tsing Hua University, Hsinchu 300044, Taiwan}
\address[2]{Institute of Astronomy Space and Earth Science, P 177, CIT Road, Scheme 7m, Kolkata 700054, India}
\address[3]{Department of Physics, National Tsing Hua University, Hsinchu 300044, Taiwan}
\address[4]{Institute of Space Engineering, National Tsing Hua University, Hsinchu 300044, Taiwan}
\address[5]{Center for Theory-Computation-Data Science Research (CTCD), National Tsing Hua University, Hsinchu 300044, Taiwan}
%\address[3]{Indian Institute of Astrophysics, Koramangala, Bengaluru, Karnataka 560034}

%\institute{Institute of Astronomy, National Tsing Hua University, Hsinchu 300044, Taiwan\\
%	  \and Institute of Astronomy Space and Earth Science, P 177, CIT Road, Scheme 7m, Kolkata 700054, India\\
%	  \and Department of Physics, National Tsing Hua University, Hsinchu 300044, Taiwan\\
%	  \email{dipakcsp@gmail.com; hkchang@mx.nthu.edu.tw}}

%\date{Received September 30, 20XX}

%==================================================================================================================================

\begin{abstract}
The Galactic transient black hole candidate MAXI J1834-021 exhibited `faint' outburst activity for approximately $10$ months following 
its discovery on February 5, 2023. We study the evolution of both the temporal (hard and soft band photon count rates, hardness ratios, and 
QPO frequencies) and spectral properties of the source using NICER data between March 7 and October 4, 2023. The outburst profile and the 
nature of QPOs suggest that the source underwent a rebrightening following the primary episode of the outburst. A monotonic evolution of 
low-frequency QPOs from higher to lower frequencies is observed during the primary episode. Both phenomenological 
(diskbb plus powerlaw) and physical (Two Component Advective Flow) model fitted spectral studies suggest that during the entire epoch, 
the source remained in harder spectral states, with a clear dominance of nonthermal emissions from the `hot' Compton cloud. Based on 
the evolution of the spectral and temporal properties, the 2023 outburst activity of MAXI J1834-021 can be classified as a `failed' outburst, 
as no softer spectral states are observed. The spectral analysis with the TCAF model also gives an estimate of the 
source mass as $12.4\pm1.7~M_\odot$.
\end{abstract}


\end{frontmatter}

\section{Introduction}

Stellar-mass black holes (BHs), especially transient black holes, are fascinating objects to study in X-rays, as they exhibit 
rapid evolution in their timing and spectral properties during their active i.e., outburst phases. Various temporal 
and spectral features, along with their evolution, are generally observable during the outbursts of transient black hole 
X-ray binaries (BH-XRBs). These timing and spectral properties, including quasi-periodic oscillations (QPOs), jets, and outflows, 
are found to be strongly correlated with each other \citep[see][]{RM06, Belloni05, Nandi12, Jana16, D21, D26a}.

The black hole continuum spectrum mainly consists of two components: a thermal multi-color disk blackbody (DBB) and a non-thermal 
emission sometimes modeled as power-law (PL). In addition to line emissions, reflection features from ionized plasma are 
sometimes observable in the BH spectrum. Spectral states are generally classified based on the relative dominance of the 
thermal and non-thermal components and the nature of QPOs \citep[see][and references therein]{Nandi12}. The two component 
advective flow (TCAF) model classifies spectral states based on the variation of the model fitted accretion flow 
parameters, accretion rate ratio (ARR), and nature of the QPOs \citep[for more details, see][]{D15, D18}.
A classical or type-I outburst of a transient black hole candidate (BHC) typically evolves through four distinct 
spectral states: low hard (LHS), hard-intermediate (HIMS), soft-intermediate (SIMS), and high soft (HSS). These states 
form a hysteresis loop, with the LHS as the two terminal states and the intermediate and soft states in between 
\citep[see][and references therein]{Nandi12}. However, in a `failed' or type-II outburst, softer states (and sometimes 
even intermediate states) are found to be missing \citep[see][and references therein]{D17}.

Although there is ongoing debate about what triggers an outburst in a transient BHC, it is generally believed that an 
outburst is initiated by a sudden enhancement of viscosity at the outer edge of the disk \citep{Ebisawa96}. 
Recently, \citet{C19} proposed that matter supplied by the companion accumulates at the pile-up radius ($X_p$), located at 
a large distance between the BH and the Lagrange point L1. During the accumulation or quiescence phase, a significant 
amount of matter gathers at $X_p$, heating up the flow on the equatorial plane until convective instability sets in and 
increases viscosity. When viscosity crosses a critical threshold, it triggers the onset of a new outburst. In a type-I 
outburst, all accumulated matter at $X_p$ is cleared, whereas in a type-II outburst, matter is only partially cleared, 
leaving behind residual material that is eventually expelled during the next type-I outburst along with freshly 
accumulated matter \citep{C19, Bhowmick21, Chatterjee22}.

Low-frequency ($0.01$–$30$ Hz) quasi-periodic oscillations (LFQPOs) are commonly observed during the hard and intermediate 
spectral states of transient BHCs. QPOs appear as peaks in the Fourier transformed power-density spectra (PDS) of light 
curves, characterized by narrow noise components, and arise due to rapid quasi-periodic variability in X-ray intensities. 
They provide crucial insights into the dynamics of accretion flows around BHs. Based on their properties  (centroid 
frequency, Q-value, rms amplitude, noise, lag, etc.), LFQPOs are classified into three types: A, B, and C \citep{Casella05}. 
Generally, the primary frequency of type-C QPOs evolves monotonically during the LHS and HIMS of both the rising and 
declining phases of an outburst. In contrast, type-B and type-A QPOs are sporadically observed in the SIMS \citep[see][]{Nandi12}. 

To understand the origin of different types of QPOs, one can find many models in literature. Some of these models are: 
Lense–Thirring precession model \citep{Stella99}, precessing hot flow model \cite{Ingram11}, accretion–ejection instability model 
\citep{Tagger99}, magnetoacoustic resonance oscillation model \citep{Titarchuk98}, shock oscillation model \citep{MSC96,RCM97}.
According to the shock oscillation model (SOM) developed by Chakrabarti and collaborators \citep{MSC96,RCM97}, LFQPOs 
originate due to oscillations of the shock. In the Two-Component Advective Flow (TCAF) solution \citep{CT95}, a hot 
Comptonizing region, known as `CENBOL,' naturally forms during the accretion process in the post-shock region. 
In the SOM framework, shock oscillations occur due to the heating and cooling effects within the CENBOL. The model 
suggests that sharp type-C QPOs arise from resonance oscillations of the shock, while type-B QPOs occur either due to 
the non-satisfaction of the Rankine-Hugoniot condition or due to a weakly resonating CENBOL. The broader type-A QPOs are 
attributed to weak oscillations of the shockless centrifugal barrier \citep[see][]{C15}. To explain the evolution of 
QPO frequencies during the rising and declining phases of transient BHCs, a time-varying form of the SOM, namely the 
propagating oscillatory shock (POS) model, was introduced by Chakrabarti and his collaborators in 2005 to study the 
evolution of QPO frequencies during the 2005 outburst of the well-known Galactic transient BHC GRO~J1655$-$40 
\citep[see][]{C08}. This POS model has been able to successfully explain evolutions of type-C QPOs in many 
transient BHCs during their rising and declining phases of outbursts \citep{C08,Nandi12,D10,D13,D17,D25,D26a}.

Accretion flow properties of black holes can be well understood through a detailed study of the spectral and temporal 
properties of the sources using physical accretion disk models. These models provide a clear picture of the flow dynamics 
of BH-XRBs during their active phases. \texttt{NthComp} \citep{Zdziarski96,Zycki99}, \texttt{kerrbb} 
\citep{Li05}, \texttt{pexrav} \citep{Magdziarz95}, \texttt{relxill} \citep[][and references therein]{Garcia14}, and 
\texttt{TCAF} \citep{CT95,D14,D15} are some of the widely used physical models. 
%The \texttt{NthComp} model is used to understand nonthermal Comptonized flux contributions and its high-energy roll-over, 
%while the \texttt{kerrbb} model is a multi-temperature blackbody model for a thin relativistic accretion disk around a 
%Kerr black hole. The \texttt{pexrav} model describes an exponentially cut-off power-law spectrum reflected from neutral 
%material, such as Fe and Ni. The \texttt{relxill} model is a relativistic accretion disk model that effectively explains 
%the reflection features observed in black holes. 
The \texttt{TCAF} model considers two types of accretion flows: a Keplerian disk (high viscosity, 
high angular momentum, geometrically thin, and optically thick) and a sub-Keplerian halo (low viscosity, low angular 
momentum, geometrically thick, and optically thin) to explain the physical processes around black holes. These physical 
models also provide an estimate of intrinsic source parameters such as mass, spin, distance, and inclination angle.

The Galactic `faint' transient BHC MAXI~J1834$-$021 was first detected by MAXI/GSC \citep{Negoro23} on February 05, 2023 
(MJD 59980). Unfortunately, it was reported nearly a month later, on 2023 March 6 (MJD 60009). Based on data from 
February 28, 2023 (MJD 60003.95) to March 1, 2023 (MJD 60004.53), \citet{Negoro23} estimated the source 
location as $(RA, Dec) = (278^\circ.634, -2^\circ.130) = (18^h34^m32^s, -02^\circ07'47'')$ (J2000) with a 90\% confidence 
level. The maximum average flux of the source during the aforementioned period was observed to be $18 \pm 4$~mCrab in the 
$4$–$10$keV MAXI/GSC band, with a $1\sigma$ error. Subsequently, MAXI~J1834-021 was monitored in multiple wavelength bands, 
including X-rays \citep[Swift, NICER, NuSTAR; see][]{Kennea23,Marino23,Homan23}, and optical \citep[LCO and Faulkes 
telescopes; see][]{Saikia23}. The source was not detected in the $15.5$~GHz band by the AMI-LA radio telescope 
\citep{Bright23}. Based on preliminary spectral and temporal analyses, including the detection of LFQPOs using NICER data, 
\citet{Homan23} confirmed the source as a black hole low-mass X-ray binary. \citet{Manca25} carried out a detailed study 
of the spectral and timing properties of the source. %mainly with the \texttt{nthComp} model. whereas in this {\it paper}, 
%to understand the accretion flow dynamics of the source during its discovery outburst, we studied the entire outburst 
%using the physical \texttt{TCAF} model. %A clearer physical picture about the evolution of the accretion flow geometry 
%during the outburst is obtained by studying the evolution of the accretion flow parameters, which are directly obtained 
%from the spectral analysis.

In this study, we analyze the evolution of both the temporal and spectral properties of the Galactic transient BHC 
MAXI~J1834$-$021 during its 2023 discovery outburst to understand the accretion flow dynamics of the source using 
archival data from NICER. By studying the variations in outburst profiles, hardness ratios, QPO frequencies, and 
spectral model-fitted parameters, we aim to gain insights into the accretion flow dynamics of the source. The spectra 
are fitted using both phenomenological (a combined disk blackbody plus power-law) and physical accretion flow (TCAF) 
models. %The spectral analysis using the \texttt{TCAF} model also provides an estimate of the probable mass of the source. 
The \textit{paper} is organized as follows: \S 2 describes the observations, data reduction, and analysis procedures. 
In \S 3, we present the results, while \S 4 discusses our findings and presents the conclusions.

\section{Observation and Data Analysis}

\subsection{Observations}

We study archival data from $98$ observations of the NICER/XTI instrument, taken roughly on a daily basis when good 
signal-to-noise ratio data are available. The data span from March 07, 2023 (MJD = 60010.01) to October 04, 2023 
(MJD = 60221.37). NICER started monitoring the source more than one month after its discovery by MAXI/GSC on 
February 05, 2023 (MJD 59980). 

\subsection{Data Reduction}  

We follow the standard data reduction procedures for the NICER satellite.
\texttt{NICER} is not an independent satellite; its X-ray Timing Instrument \citep[XTI;][]{Gendreau12} is attached as 
an external payload to the International Space Station. It operates in the energy range of 0.2--12~keV with a time 
resolution of $\sim0.1$~${\mu}$s and a spectral resolution of $\sim 85$~eV at 1~keV. For data analysis, we use the online 
platform SciServer\footnote{\url{https://www.sciserver.org}} with HEASARC's latest HEASoft package, version 6.34. 
The Level 1 data files are processed with the \texttt{nicerl2} script in the latest CALDB environment to obtain fully 
calibrated Level 2 event files, which are then screened for non-X-ray events or bad data times. Light curves with $10$~s, $1$~s 
and $0.01$~s time bins in the energy bands $0.5$–$3$~keV, $3$–$10$~keV, and $0.5$–$10$~keV are then extracted using the 
task \texttt{nicerl3-lc}. To obtain spectra in the default energy band, we use the task \texttt{nicerl3-spect} with 
{\fontfamily{qcr}\selectfont SCORPEON}\footnote{\url{https://heasarc.gsfc.nasa.gov/docs/nicer/analysis_threads/scorpeon-xspec/}} 
background model.

\subsection{Data Analysis}

We use NICER light curves to determine the average count rates per observation in different energy bands.
%The average count rate ($R_{\rm obs}$) and its uncertainty ($\sigma_{R}$) are calculated from the total number of counts ($N_{\rm tot}$) and the total exposure time ($T_{\rm exp}$) for each observation using formulae: $R_{\rm obs} = \frac{N_{\rm tot}}{T_{\rm exp}}$, and $\sigma_{R} = \frac{\sqrt{N_{\rm tot}}}{T_{\rm exp}}$.
To study power density spectra (PDS), we analyze $0.01$s time-binned light curves in the $0.5-10$~keV band. 
To determine the parameters (centroid frequency, full width at half maximum [FWHM], and power) of the QPOs,
white-noise subtracted PDS (using the \texttt{`powspec~norm=-2'} task) are fitted with a 
Lorentzian model using the \texttt{XRONOS} package of HEASoft. The Lorentzian fit parameters are then used to calculate 
the Q-value ($\nu/{\rm FWHM}$), fractional rms (\% rms $= 100~\sqrt{\pi \times \text{power} \times \text{FWHM}/2}$), and 
the significance ($\sigma$, defined as the ratio of the normalization to its negative error) of the QPOs.

The $0.5-10$~keV NICER spectra are fitted with a combination of thermal disk blackbody (DBB) and power-law (PL) models, 
and also with the \texttt{TCAF} model in \texttt{XSPEC} \citep{Arnaud96}. For spectral analysis with the physical TCAF 
model, we used the model fit file (v0.3.2) as a local additive table model \citep[for more details about the generation 
and parameter space of the fits file, see][]{D15,D18}. To account for interstellar absorption, we employed the 
\texttt{TBabs} model with the hydrogen column density fixed at $N_{\rm H}=9\times10^{21}~{\rm cm^{-2}}$. This value 
corresponds to the mean best-fit hydrogen column density obtained from the physical TCAF model fits when $N_{\rm H}$ 
was allowed to vary freely. Furthermore, to account for residual instrumental features in the \textit{NICER} spectra, 
we included an additional \texttt{edge} component with the absorption edge energy at $\sim0.8$~keV. 
Note that the low signal-to-noise ratio and background contamination above $\sim6$--$7$~keV restricted the spectral 
analysis to only $58$ of the total $98$ observations. 
%In these leftover observations, the spectral fits would not be reliable, particularly for the TCAF model, because 
%the high-energy nonthermal emission from the Compton cloud, which is essential for constraining the model parameters, 
%is not adequately detected.}
%(Contact to \href{https://csp.res.in/index.php?name=bVdaWERUZThCNUdqYktvSGk4VnREZz09OjoxMjM0NTY3ODkxMjM0NTY3}
%{Prof. S. K. Chakrabarti}) via email: sandip@csp.res.in). %\footnote{Contact Prof. S. K. Chakrabarti}.

\begin{figure}%[!h]
  \centering
    \includegraphics[angle=0,width=9.0cm,keepaspectratio=true]{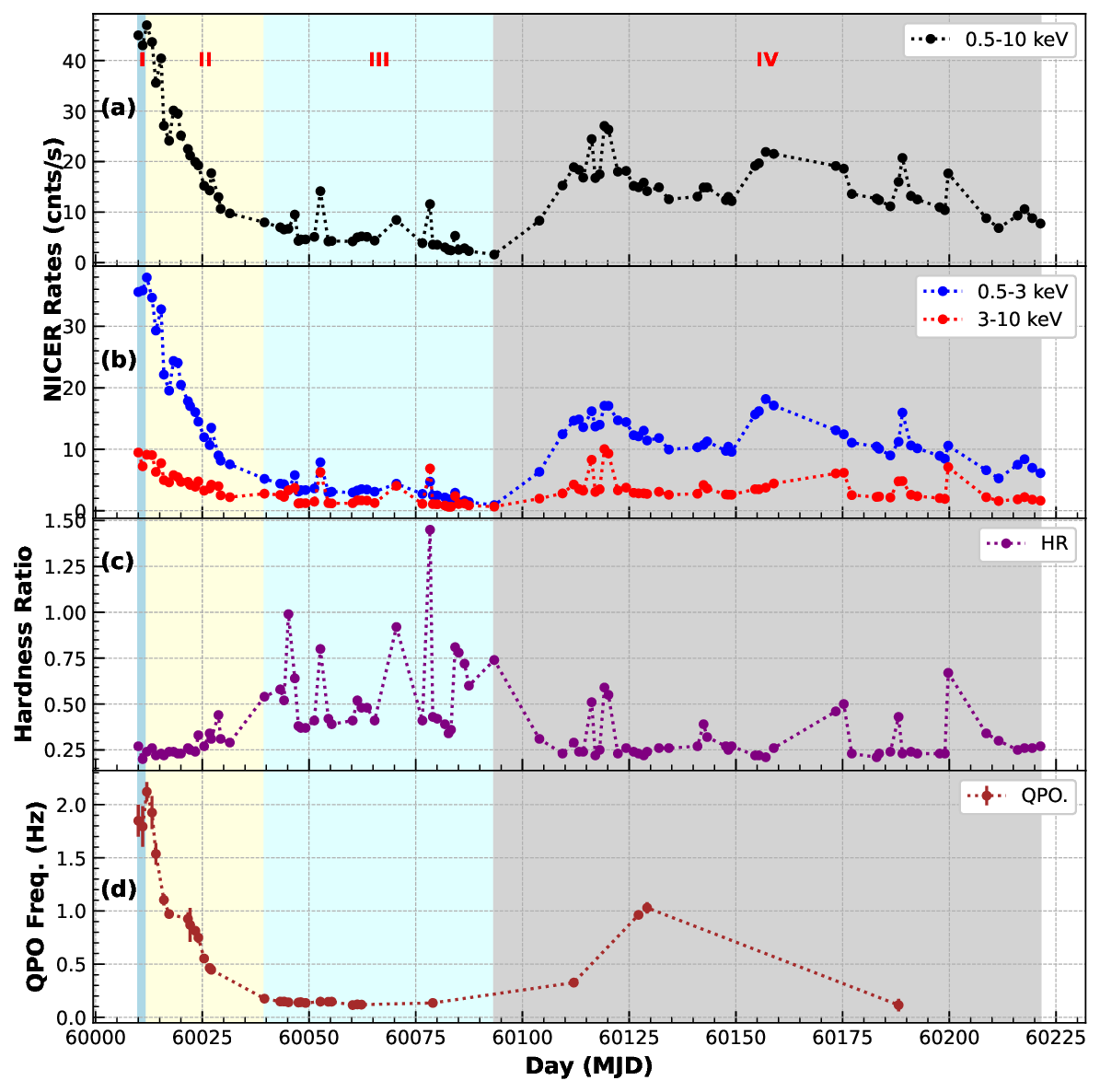}
\caption{(a-b) Variation of NICER count rates in the soft X-ray (SXR; $0.5$–$3$~keV), hard X-ray (HXR; $3$–$10$~keV),  
        and total X-ray (TXR; $0.5$–$10$~keV) energy bands, shown in the top two panels.  
        (c) Evolution of the hardness ratio (HR = HXR/SXR) is shown in the next panel.  
        (d) In the bottom panel, the variation of the observed QPOs is shown. The shaded regions mark different phases 
        of the outburst profile.  Phases I–III correspond to the primary outburst phase, while Phase IV belongs to 
        the secondary rebrightening phase.}
\label{fig_lchr}
\end{figure}

%\newpage
\section{Results}

\subsection{Outburst Profile}

%We used $10$~s time-binned light curves from the 
We use archival data from NICER/XTI instrument over a period of $\sim 8$ months (from March 07 to 
October 04, 2023; i.e., MJD 60010.01–60221.37) to study the outburst profile of the discovery outburst activity of the 
Galactic transient BHC MAXI J1834-021. Panels (a–b) of Fig.~\ref{fig_lchr} show the evolution of the average count rates in 
the total (TXR; $0.5$–$10$~keV), soft (SXR; $0.5$–$3$~keV), and hard (HXR; $3$–$10$~keV) X-ray bands. From the outburst 
profiles, it appears that we missed the initial rise in flux, which is typically observed in the canonical outburst of a 
transient BHC. This is because NICER began monitoring the source $\sim 32$~days after its discovery.

Based on the evolution of the outburst profiles, hardness ratios, and QPOs, the 2023 activity of the transient BHC 
MAXI J1834-021 can be classified into four phases. In Phase I (March 07–09, 2023; MJD 60010.01–60012), we observe a 
rise in the SXR and TXR rates, with peaks in both the rates occurring on March 09, 2023 (MJD 60012). In Phase II 
(March 09–April 05, 2023; MJD 60012–60039.57), both TXR and SXR exhibit a decreasing trend. The HXR also follows a 
decreasing trend in both Phases I \& II, except for a sudden glitch on March 24, 2023 (MJD 60027.11). In Phase III 
(April 05–May 29, 2023; MJD 60039.57–60093.39), all band rates remain roughly constant in a low-luminosity phase. 
After this, in Phase IV (May 29–October 04, 2023; MJD 60093.39–60221.37), we observe a rebrightening episode 
with multiple rises and drops in count rates. However, by the end of our observations, all band count rates had not 
reached their lowest values observed in Phase III, suggesting that the outburst might have continued beyond our dataset.

Overall, the entire 2023 activity of the source can be classified into two epochs consisting primary and reflare 
episodes of the outburst. Phase I corresponds to the rising phase, while Phases II \& III represent the declining phases 
of the primary episode. Phase IV constitutes the rebrightening or reflare episode of the outburst. 
%\textbf{According to the rebrightening classification method proposed by \citet{Zhang19}, the 2023 activity of MAXI J1834-021 
%contains one reflare or rebrightening episode following the primary episode of the outburst.} 

\begin{figure}%[!h]
  \centering
    \includegraphics[angle=0,width=9.0cm,keepaspectratio=true]{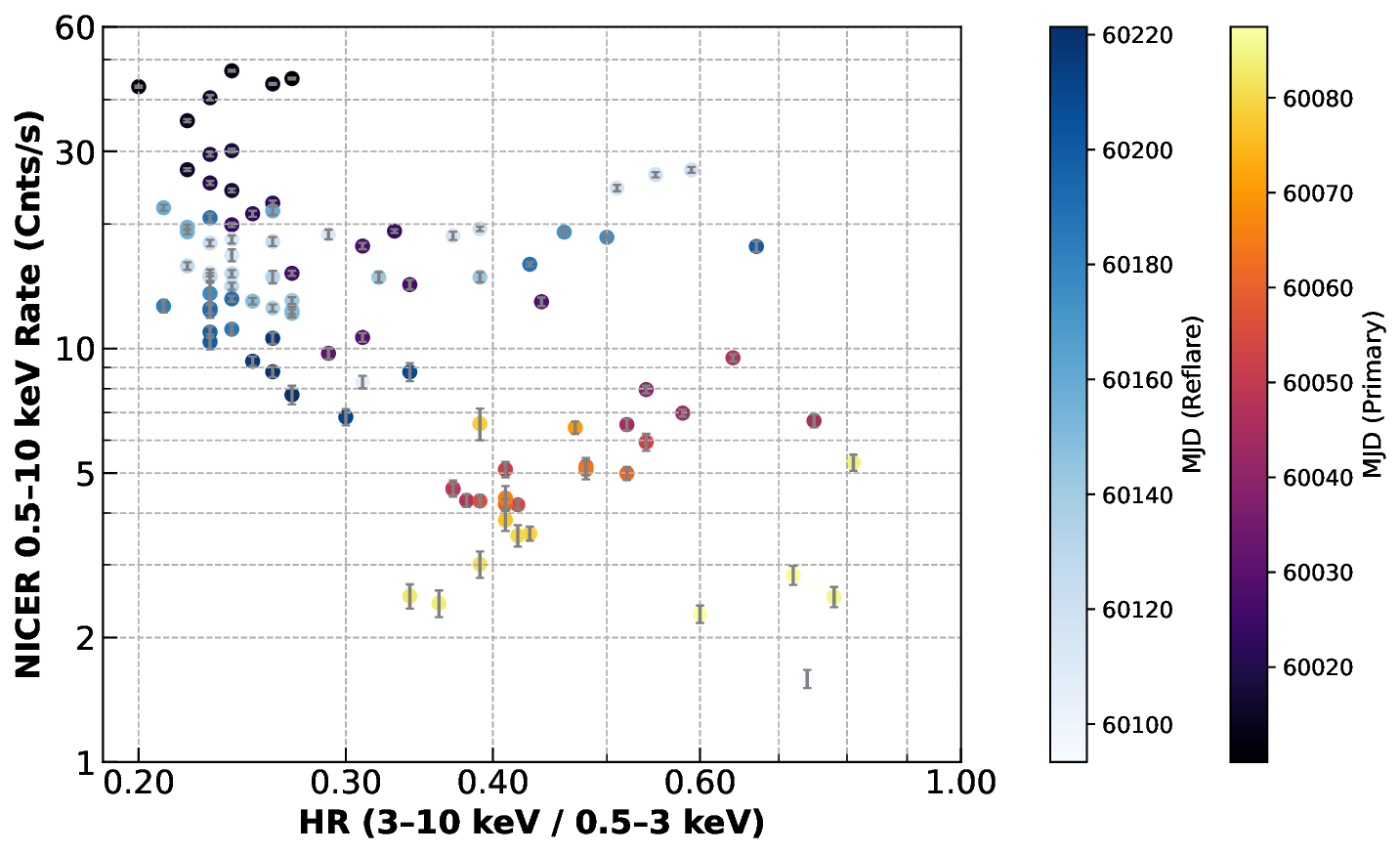}
    \caption{Intensity vs. Color-Color or Hardness Intensity Diagram (HID) in log-log scale. Colors in the two bars present MJDs from primary 
	and reflare episodes during the 2023 outburst of MAXI J1834-021.}
\label{fig_hid}
\end{figure}

\subsection{Hardness Ratio (HR) and Hardness-Intensity Diagram (HID)}

The variation of the hardness ratio (HR) or X-ray color-color ratio is shown in Fig.~\ref{fig_lchr}(c). We define HR as the 
ratio of the $3$–$10$~keV HXR count rate to the $0.5$–$3$~keV SXR count rate. During the short duration of Phase I, HR is 
found to decrease, whereas in Phase II, it shows an increasing trend. The decreasing and increasing trends observed in 
Phases I \& II, respectively, resemble the HR evolution in the rising and declining HIMS \citep[for more details, see][]{Nandi12}. 
In Phase III, higher HR values are observed, which is characteristic of LHS. In Phase IV, HR values remain at a lower level. 
%\textbf{with the characteristic of the unclassified LHS and HIMS.}

We show the variation of the intensity vs. color--color i.e., hardness--intensity diagram (HID) in Fig.~\ref{fig_hid}. 
In the HID, two color bars mark the primary and reflare episodes of the outburst, with colors corresponding to the observation 
dates (MJD). During the primary episode of the outburst, due to opposite trends in intensity and HR, the HID data points shift 
toward the lower right as the outburst progresses. This behavior is typically observed during the decay phase of canonical 
transient BH-XRBs, reflecting a progressively harder spectral state.
In contrast, during the reflare episode, the X-ray intensity is observed to vary within a narrow range despite variations in HR. 
Overall, the nature of the HID evolution also suggests that the outburst is of type-II or `failed', as the canonical `q'-shaped 
HID typically observed in classical outbursts of transient BH-XRBs is not seen.  

\begin{figure}%[!h]
  \centering
    \includegraphics[angle=0,width=9.0cm,keepaspectratio=true]{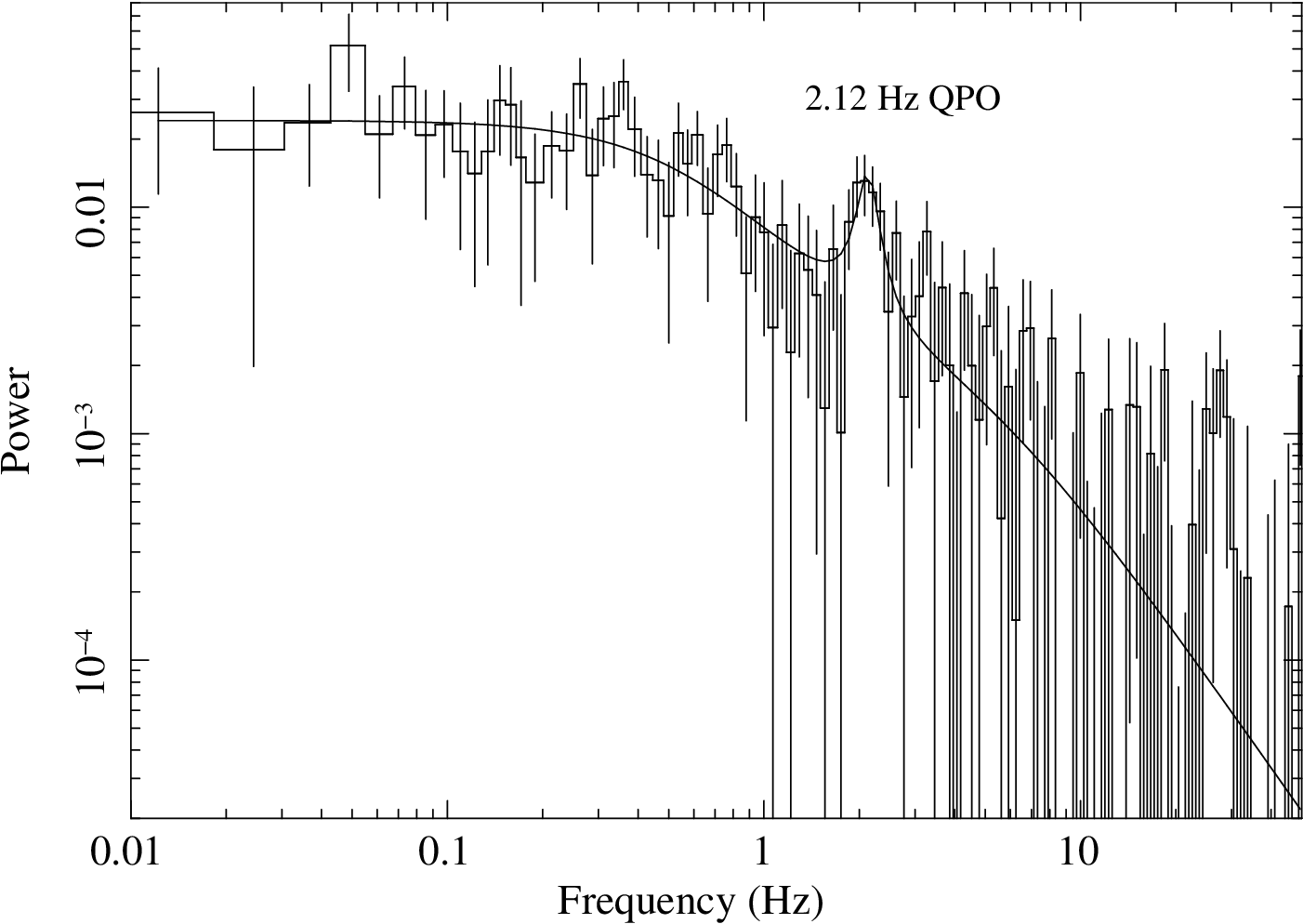}
    \caption{Power density spectrum (PDS) of the $0.01$~s time-binned $0.5$–$10$~keV NICER light curve 
	from March 09, 2023 (MJD 60012). The Lorentzian model fit reveals a prominent QPO at $2.12$~Hz.}
\label{fig_pds}
\end{figure}

\subsection{Low frequency QPOs (LFQPOs)}

LFQPOs are important temporal features commonly observed in transient BHCs. They generally also show evolution 
with spectral states during both the rising and declining phases of an outburst. During the 2023 outburst of MAXI~J1834-021, 
QPOs are detected in $32$ out of a total of $98$ NICER observations (see Table~\ref{table_rate_qpo}).
In Phase I, we observe an increasing trend in QPO frequencies from $1.84$ to $2.12$~Hz. Subsequently, in Phases II \& III, 
QPO frequencies monotonically decrease from $2.12$~Hz to $0.19$~Hz. This pattern of increasing and decreasing QPO frequencies 
is commonly observed in the rising and declining harder states of transient BHCs respectively \citep[see][]{C08,Nandi12, D13}. 
In Phase IV, QPO frequencies exhibit both increasing and decreasing trends. The highest QPO frequency during the entire outburst 
is observed at $2.12$~Hz on March 09, 2023 (MJD 60012), while in Phase IV, the maximum QPO frequency is recorded at $1.03$~Hz 
on July 04, 2023 (MJD 60129.20). The power density spectrum (PDS) of the QPO, fitted with a combination of two 
zero-centered and one QPO-centric Lorentzian models, is shown in Fig.~\ref{fig_pds}.

\begin{figure}[t]
  \centering
  \includegraphics[angle=0,width=8.8cm,keepaspectratio=true]{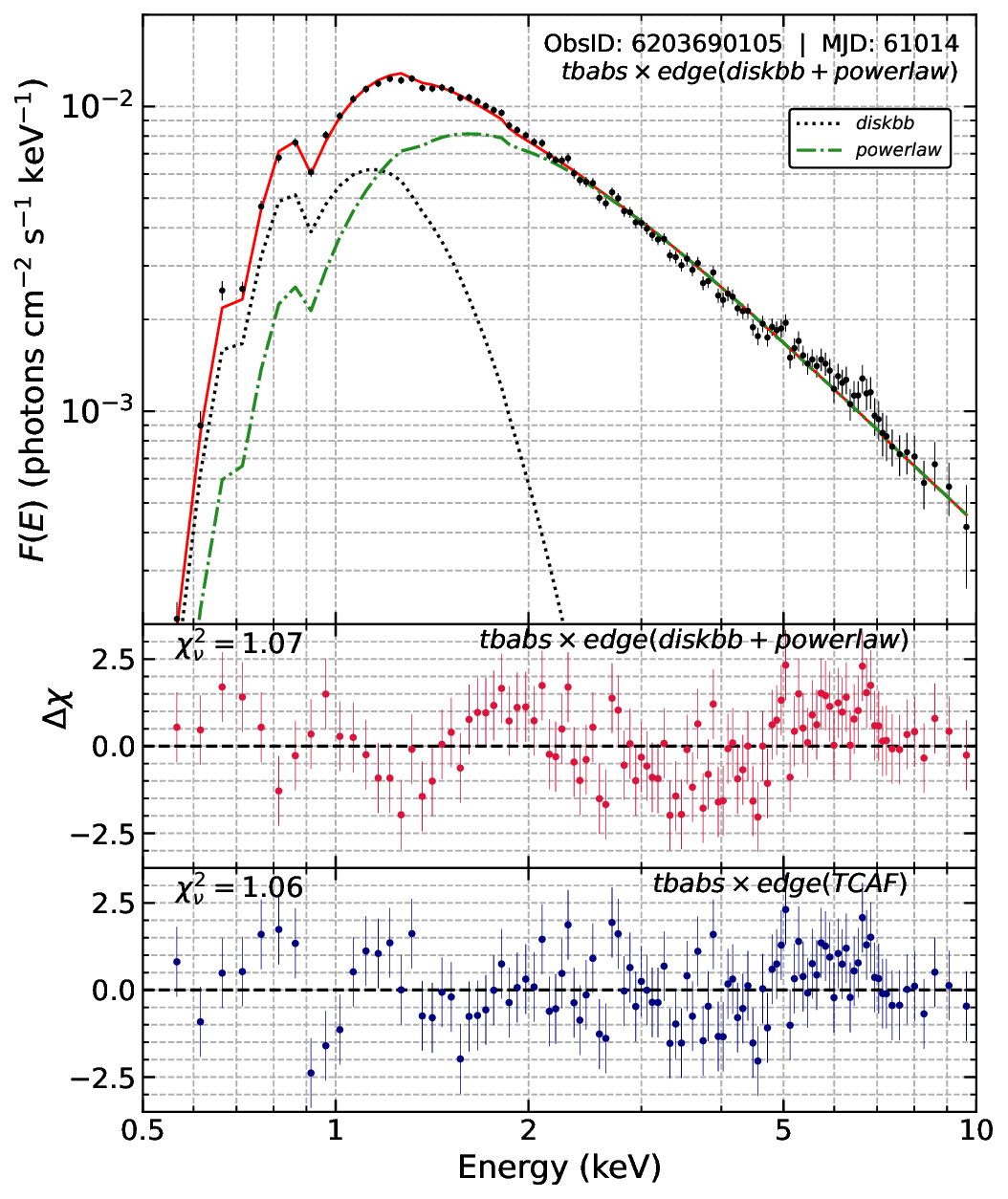}
	\caption{NICER spectrum of observation ID: 6203690105 (from March 11, 2023; MJD=60014.13) fitted with two sets of 
    models: $(i)$ $tbabs\times smedge(diskbb + powerlaw)$ (red dashed line), and 
    $(ii)$ $tbabs \times smedge \times TCAF$ (blue dash-dotted line). 
    The purple and brown colored solid curves are for model set1 fitted disk blackbody and powerlaw components.} 
\label{fig_sfit}
\end{figure}

%\newpage
\subsection{Spectral Study using Phenomenological Model}

For a better understanding of the evolution of accretion flow dynamics in MAXI~J1834-021 during its 2023 outburst, 
we studied the spectral nature of the source using either a combined disk blackbody plus power-law 
model ($TBabs\times edge(diskbb+powerlaw)$) or only a power-law model ($TBabs\times edge\times powerlaw$) for 
NICER spectra in the $0.5$–$10$~keV band with a fixed  $N_{\rm H}=9\times10^{21}~{\rm cm^{-2}}$. Out of a total of 
$98$ NICER observations, we used $58$ for spectral analysis (see Table~\ref{table_spec}). The same 
spectra are further studied with the physical TCAF model (see sub-section 3.5). 

A sample spectrum from March 11, 2023 (MJD = 60014.13), fitted with both the combined DBB plus PL model, and the TCAF 
model, is shown in Fig.~\ref{fig_sfit}. The evolution of the DBB and PL components indicates that the thermal 
disk emission dominates below $2$~keV, while the higher-energy spectrum is governed by the non-thermal PL component. 
The residuals and reduced $\chi^2$ ($\chi^2_{\nu}$) values suggest a marginally improved fit with the physical TCAF model.
%The omitted observations had low signal-to-noise ratios, or we were unable to generate spectral files (PHA, RMF, and ARF).

%For the best model fits, based on the reduced $\chi^2$ ($\simeq 1$), the hydrogen column density in the interstellar 
%absorption model {\fontfamily{qcr}\selectfont TBabs} is obtained in the range of $(0.42$–$0.87)\times 10^{22}$~atoms/cm$^2$. 
%The {\fontfamily{qcr}\selectfont edge} model, with an edge energy of $\sim 0.81$~keV, is also used to account for 
%instrumental features in the NICER spectra. 

\begin{figure}[!t]
\vskip -0.3cm
  \centering
    \includegraphics[angle=0,width=9.2cm,keepaspectratio=true]{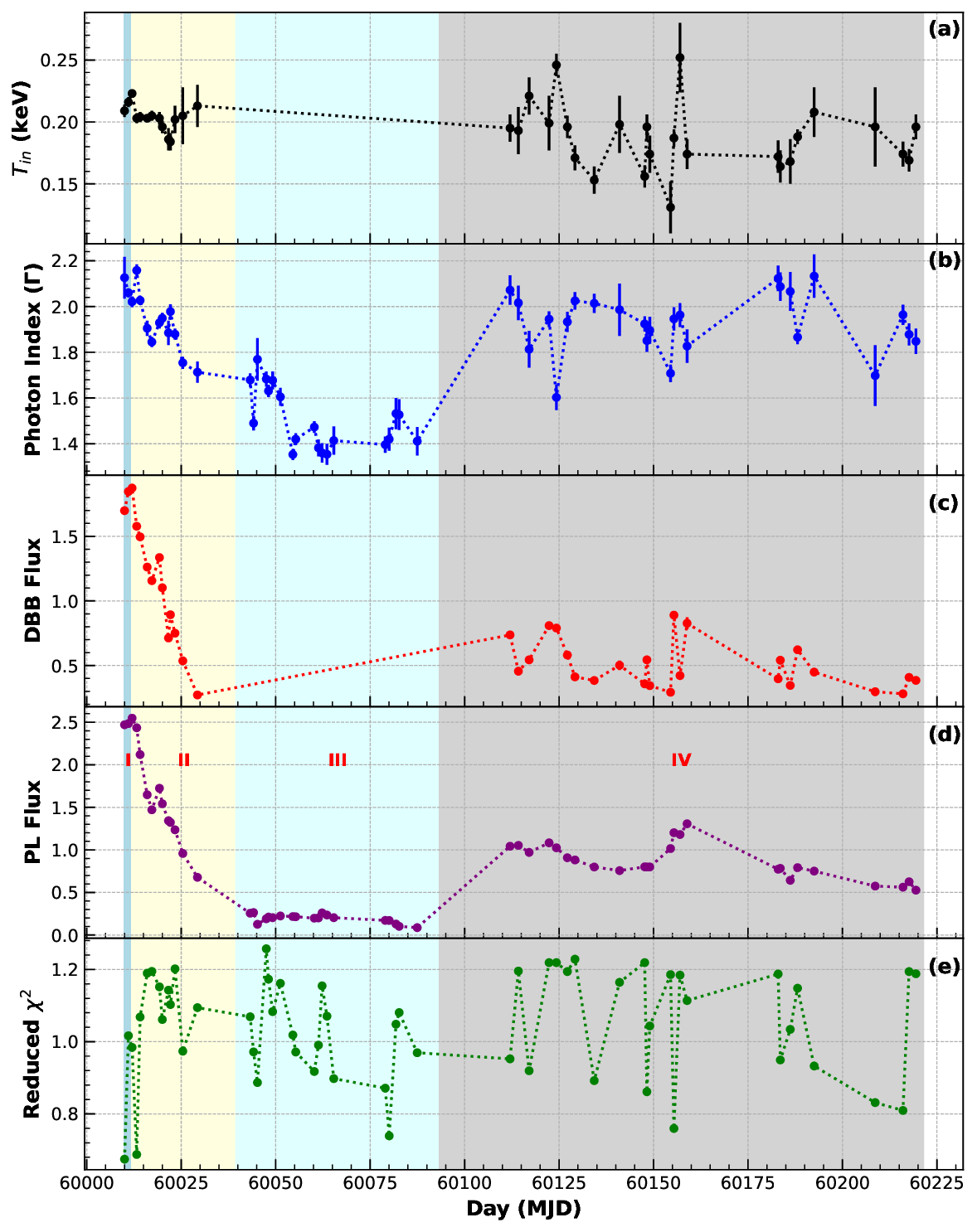}
        \caption{Variation of the spectral parameters obtained from the combined disk blackbody (DBB) and power-law (PL) 
        model fit: (a) disk temperature ($T_{\rm in}$ in keV), (b) power-law photon index ($\Gamma$), 
        (c-d) fluxes of the model components (both in units of 
        $10^{-10}\text{ergs}\text{cm}^{-2}~\text{s}^{-1}$), and (e) model fitted reduced $\chi^2$.
        The shaded regions indicate different Phases of the outburst.} 
\label{fig_dbbpl}
\end{figure}

In Fig.~\ref{fig_dbbpl}, we show the variation of the DBB temperature ($T_{\rm in}$ in keV), PL photon index ($\Gamma$), 
and DBB and PL model component fluxes in the fitted spectral energy range of $0.5$–$10$~keV. 
The presence of a cooler disk component, with a temperature $T_{\rm in}$ in the range of $0.13$--$0.25$~keV, is observed 
in the observations where the combined DBB plus PL model is used.
No distinct evolutionary trend in $T_{\rm in}$ is observed during 
the outburst. However, we find the absence of the DBB component in Phase III, where the spectra are fitted only with 
the PL model. The PL photon index $\Gamma$ is observed to vary between $1.35$ and $2.16$, indicating that the source 
remained in harder states, i.e., in the LHS and HIMS.

The DBB (F$_{\rm DBB}$) and PL (F$_{\rm PL}$) fluxes show an increasing trend over a short period in Phase I. Then, in Phase II, 
both fluxes decline rapidly. The peak of both the thermal DBB and nonthermal PL fluxes is observed on the transition day (MJD = 60012) 
between Phases I and II. Lower values of the PL flux are observed in Phase III, where the disk component is absent. 
In Phase IV, both DBB and PL fluxes show variations that are roughly consistent with the outburst profiles (in the SXR 
and HXR bands) shown in Fig.~\ref{fig_lchr}. Throughout the entire outburst, a clear dominance of the nonthermal PL model 
component is observed.

A clear distinction is observed in the evolution of $\Gamma$ and the model-component fluxes between the primary 
and reflare episodes of the outburst. During the primary episode, these parameters generally exhibit a decreasing trend, 
except during the initial short-lived Phase~I. In contrast, during the reflare episode, neither PL photon index $\Gamma$ 
nor the disk black body flux $F_{\rm DBB}$ shows any systematic trend, similar to the behavior of disk temperature 
$T_{\rm in}$. However in the Phase-IV, the PL flux $F_{\rm PL}$ exhibits a more complex evolution: it remains nearly constant 
during MJD~60112--60124, decreases until MJD~60134, subsequently increases until MJD~60159, and finally decreases again. 
%Overall, the primary and reflare episodes are characterized by variations in $T_{\rm in}$ over the ranges of $0.18$--$0.22$~keV 
%and $0.13$--$0.25$~keV, respectively; $\Gamma$ over $1.35$--$2.16$ and $1.6$--$2.1$, respectively; $F_{\rm DBB}$ over 
%$(0.27$--$1.87)\times10^{-10}$ and $(0.28$--$1.89)\times10^{-10}$~erg~cm$^{-2}$~s$^{-1}$, respectively; and $F_{\rm PL}$ 
%over $(0.09$--$2.55)\times10^{-10}$ and $(0.53$--$1.31)\times10^{-10}$~erg~cm$^{-2}$~s$^{-1}$, respectively.

\subsection{Spectral Study using Physical Model}

NICER spectra are further studied with the physical TCAF model, which provides deeper insights into the accretion flow 
dynamics of the source. The TCAF model simultaneously accounts for both types of flux contributions: thermal
emission from the Keplerian disk and nonthermal Comptonized emission from the hot corona or CENBOL. Therefore, spectral 
fitting with the TCAF model using NICER/XTI or similar satellite instruments, even within the limited energy band of around 
$0.5$--$10$~keV, can provide a more physically consistent estimation of the relative thermal and nonthermal contributions 
compared to a combination of phenomenological models, where separate components are used to represent different emission processes.
Each spectral fit with the TCAF model estimates physical flow parameters such as the Keplerian disk accretion rate ($\dot{m}_d$ 
in Eddington rate $\dot{M}_{\rm Edd}$, where $\dot{M}_{\rm Edd} \simeq 1.44 \times 10^{17}~(M_{\rm BH}/M_\odot)$~g\,s$^{-1}$), 
sub-Keplerian halo accretion rate ($\dot{m}_h$ in $\dot{M}_{\rm Edd}$), shock location ($X_s$ in Schwarzschild radius, 
$r_s = 2GM_{\rm BH}/c^2$), and shock compression ratio ($R=\rho_+/\rho_-$, where $\rho_+$ and $\rho_-$ are the post- and pre-shock 
matter densities, respectively). This model also provides an estimate of the black hole mass ($M_{\rm BH}$) directly from spectral 
fits if the mass of the source is unknown and is kept as a free parameter while fitting the BH spectrum. 

\begin{figure}[!t]
\vskip -0.3cm
  \centering
    \includegraphics[angle=0,width=9.2cm,keepaspectratio=true]{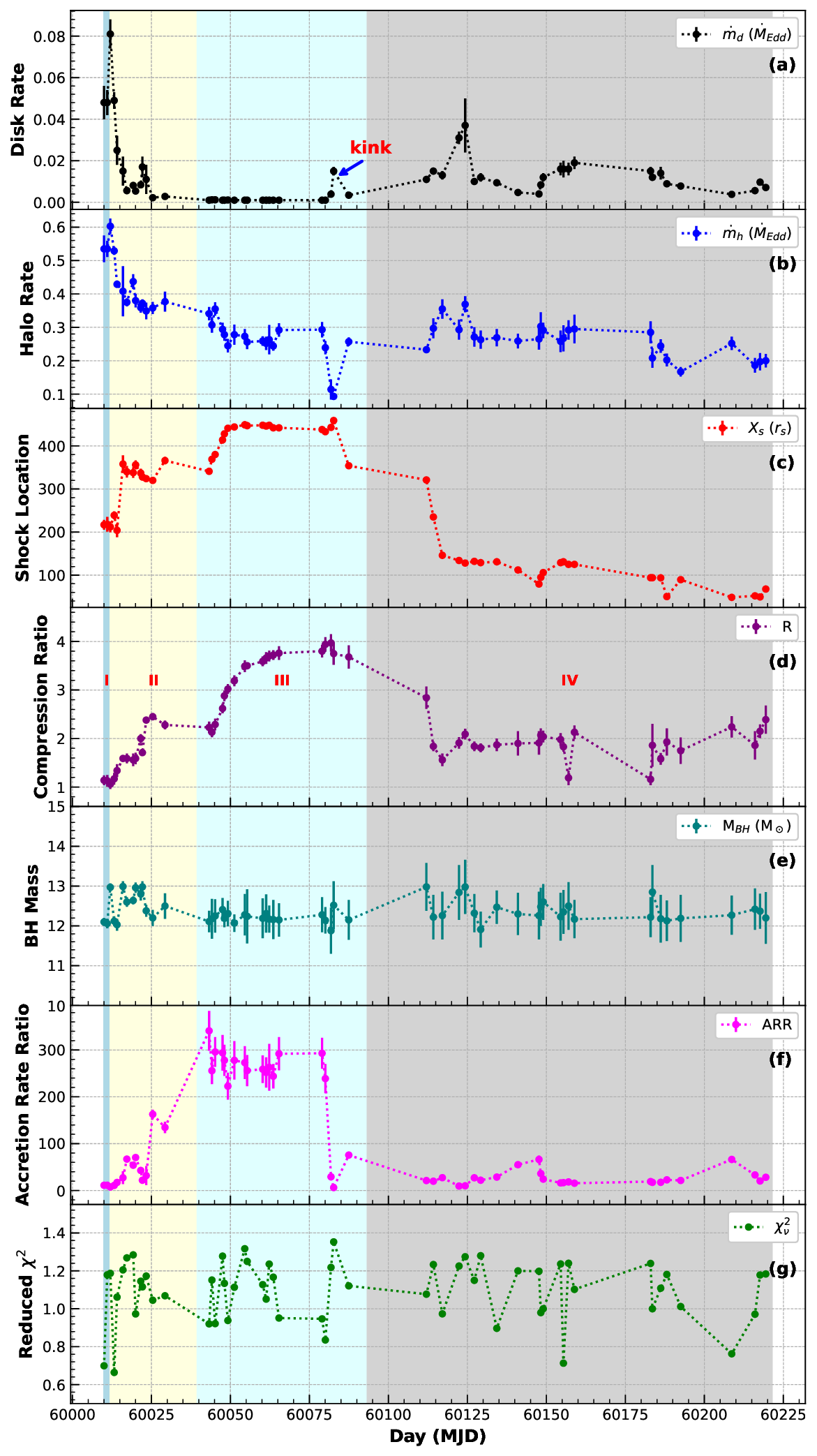}
        \caption{Variation of the TCAF model fitted spectral parameters: (a) Keplerian disk accretion rate ($\dot{m}_d$ 
        in $\dot{M}_{\rm Edd}$), (b) sub-Keplerian halo accretion rate ($\dot{m}_h$ in $\dot{M}_{\rm Edd}$), (c) Shock 
        Location ($X_s$ in $r_s$), (d) compression ratio (R), (e) BH Mass (M$_{\rm BH}$ in M$_\odot$), (f) the ratio between 
	halo to disk accretion rates as ARR are shown. In the bottom panel (g), variation of the model fitted reduced $\chi^2$ 
	is shown. The shaded regions indicate different phases of the outburst profile. Note, here `kink' on MJD=60082.68, 
	indicates possible triggering of the rise in viscosity at pile-up radius $X_p$, which might causes rebrightening 
	episode of the outburst (Phase-IV) after the primary episode of the outburst (combining Phases I-III).}
\label{fig_tcaf}
\end{figure}

The evolution of the TCAF model-fitted flow parameters ($\dot{m}_d$, $\dot{m}_h$, $X_s$, $R$), and derived parameter, 
namely the accretion rate ratio ($ARR = \dot{m}_h/\dot{m}_d$), is shown in Fig.~\ref{fig_tcaf}. Similar to the combined 
DBB plus PL model fits, we also observe peaks in both the disk ($0.081$~$\dot{M}_{\rm Edd}$) and halo 
($0.603$~$\dot{M}_{\rm Edd}$) accretion rates on MJD = 60012, the day when we observed the highest frequency of the 
evolving QPO ($\nu_{\rm QPO} = 2.12$~Hz). After that, both rates exhibit a decreasing trend. The halo accretion rate 
decreases monotonically (from $0.603$ to $0.093$ $\dot{M}_{\rm Edd}$) until 
the end of Phase III, while the disk accretion rate rapidly decreases to its lowest value ($0.001~\dot{M}_{\rm Edd}$). 
Subsequently, in phase IV, i.e., during the rebrightening episode of the outburst, both rates evolve in a manner 
roughly consistent with the outburst profiles shown in Fig.~\ref{fig_lchr}. 
A `kink' in the disk accretion rate is observed on May 18, 2023 (MJD = 60082.68), which may indicate the triggering 
of a fresh supply of matter from the pile-up radius $X_p$, leading to the reflare.
In this particular observation, the increase in $\Gamma$ also indicates a local spectral softening within the LHS.

Throughout our analysis, $X_s$ is found to vary between $48$ and $459r_s$, while $R$ varies between $1.06$ and $3.97$. 
In Phase I, both $X_s$ and $R$ decrease slightly. In Phases II and III (up to the `kink' day), a receding shock with 
an overall increasing trend is observed. On the `kink' day, the shock reaches its extreme location ($X_s = 459~r_s$) 
with nearly its highest observed strength ($R = 3.75$). After this `kink' day, an inward-moving shock with decreasing 
strength is observed. This pattern in $X_s$ and $R$ is typically seen during the rising phase of an outburst.
%Note: here $X_s=460$, and $R=1.05$ are the limiting values of the TCAF model fits file. This means that there might
%be a possible rise and decrease of the shock location and compression ratios. 

The evolution of the accretion rate ratio ARR is shown in Fig.~\ref{fig_tcaf}(f). The ARR increases until MJD~60043.26 
and then remains approximately constant at a high level until about two days before the `kink' day, indicating the higher supply 
of the low viscous and low angular momentum sub-Keplerian halo matter over the high viscous and high angular momentum Keplerian 
disk accretion flow. Thereafter, the ARR decreases rapidly, reaching its minimum on the `kink'' day. This sharp decline indicates 
a temporary increase in the relative contribution of the Keplerian disk accretion rate. Subsequently, the ARR remains within a 
relatively narrow low-value range in Phase-IV. Overall, the ARR varies between $\sim6$ and $\sim341$ throughout the 2023 
outburst of MAXI~J1834-021.
%These high ARR values suggest that, throughout the outburst, the flow was highly dominated by the sub-Keplerian halo component, which also evident from Fig.~\ref{fig_tcaf}(b).

Similar to the evolution of the DBB+PL model-fitted parameters and fluxes, the TCAF model-fitted parameters 
($\dot{m}_d$, $\dot{m}_h$, $X_s$, and $R$), together with the derived ARR, exhibit distinctly different evolutionary 
behaviors during the primary and reflare episodes of the outburst. During the primary episode, the Keplerian disk 
accretion rate and sub-Keplerian halo accretion rate generally decrease, whereas the shock location, compression ratio, 
and ARR show an overall increasing trend. The only exception is the initial short-lived Phase~I, during which the opposite 
behavior is observed in this primary episode. In contrast, during the reflare episode, $\dot{m}_d$, $\dot{m}_h$, $R$, 
and ARR remain approximately constant, apart from a few short-term fluctuations, while $X_s$ exhibits a gradual decreasing trend. 
%Overall, the primary and reflare episodes are characterized by $\dot{m}_d$ values in the ranges of $0.001$--$0.081$ 
%and $0.004$--$0.037~\dot{M}_{\rm Edd}$, respectively; $\dot{m}_h$ values of $0.003$--$0.603$ and $0.167$--$0.369~\dot{M}_{\rm Edd}$, 
%respectively; shock locations ($X_s$) of $204$--$459~r_s$ and $48$--$321~r_s$, respectively; compression ratios ($R$) of 
%$1.06$--$3.97$ and $1.16$--$2.84$, respectively; and ARR values of $6$--$341$ and $9$--$66$, respectively.

\subsection{Estimation of Mass of the Source}
The mass of the BH, $M_{\rm BH}$ is a crucial input parameter in the TCAF model. If the mass of a black hole is not 
well constrained (e.g., not estimated dynamically), it can be inferred by keeping $M_{\rm BH}$ as a free parameter 
while fitting the spectra with the physical TCAF model. In the past, this model has been successfully used to estimate 
the masses of several black hole candidates \citep[see, e.g.,][]{Molla16, Molla17, D17, Shang19, Nath24}. 
Since the $M_{\rm BH}$ of MAXI~J1834-021 is not known from dynamical methods, we kept it as a free parameter while 
fitting the spectra. The TCAF model fitted $M_{\rm BH}$ is found to be in the range $\sim11.9$--$13~M_\odot$, with an average 
value of $12.4~M_\odot$ (Fig.~\ref{fig_tcaf}e). The standard deviation ($\sigma$) of the estimated mass values is $1.7~M_\odot$; thus, 
we infer the probable mass of the BH to be $12.4 \pm 1.7~M_\odot$.
%\textbf{The weighted mean and weighted standard deviation of the estimated BH mass are found to be $12.48~M_\odot$ and $0.34~M_\odot$, respectively. Based on these estimates, we infer the probable mass of the black hole to be $12.5 \pm 0.3~M_\odot$.}
In \citet{DC26}, we presented complementary spectral and temporal methods that further support this estimation.

\begin{figure}[!t]
\vskip -0.3cm
  \centering
    \includegraphics[angle=0,width=9.0cm,keepaspectratio=true]{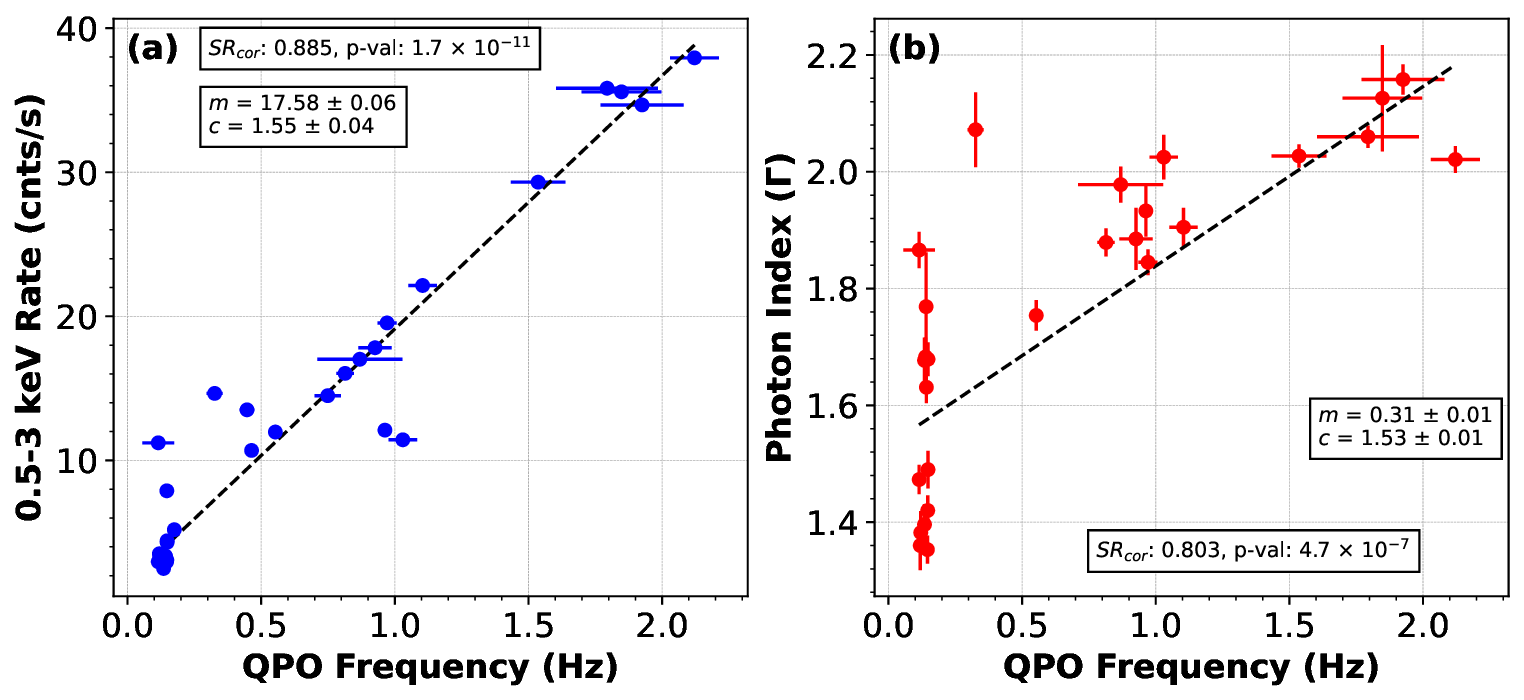}
        \caption{The variation of the QPO frequency ($\nu_{\rm QPO}$) with (a) the soft X-ray (SXR; $0.5$--$3$~keV) count rate 
	and (b) the photon index ($\Gamma$) is shown to study the correlation of $\nu_{\rm QPO}$ with the SXR and $\Gamma$. 
    The dashed black lines show a linear ($y = mx + c$) trend in the evolutions of $\nu_{\rm QPO}$ with SXR and $\Gamma$. 
    The Spearman rank correlation coefficient ($SR_{\rm cor}$), along with its $p$-value, is shown in the inset.
    }
\label{fig_corr_qpohr}
\end{figure}

\subsection{Correlation of Temporal and Spectral Properties}

The correlation of different spectral and timing parameters allows us to understand the relationships between them. 
Here, we use the Spearman rank correlation method, which is one of the fundamental statistical tools widely used in 
scientific research. The correlation coefficients of the method provide insight into patterns and dependencies 
within the data.

\subsubsection{Correlation of QPO Frequency with SXR and Photon Index}

It is evident from Fig.~\ref{fig_lchr} that the evolution of the QPO frequency ($\nu_{\rm QPO}$) follows a trend 
roughly similar to that of the SXR in the $0.5$--$3$~keV NICER band. To confirm this statistically, 
we study the correlation between $\nu_{\rm QPO}$ and the SXR using the Spearman rank correlation method 
(see Fig.~\ref{fig_corr_qpohr}a). This analysis reveals a strong positive correlation, with a correlation coefficient 
of $SR_{\rm cor} = 0.885$ and a p-value of $1.7 \times 10^{-11}$. Furthermore, a linear fit 
%with $y=mx+c$ (where $m$ is the slope and $c$ is the intercept) 
also confirms a statistically acceptable correlation between the SXR and $\nu_{\rm QPO}$. % ($\chi^2_{\nu}=0.63$).
%Due to random variations, the linear fit with $\chi^2_{\nu}>6$ is obtained for $\Gamma$ versus $\nu_{\rm QPO}$.}

We also studied the correlation between the observed QPO frequency  and PL photon index (i.e., $\nu_{\rm QPO}$ vs. $\Gamma$), 
obtained from the combined DBB plus PL model or from only PL model fitted spectra (see Fig.~\ref{fig_corr_qpohr}b). 
The Spearman rank correlation coefficient $SR_{\text{cor}} = 0.803$ with a p-value of $4.7 \times 10^{-7}$ suggests 
a strong positive correlation between $\nu_{\text{QPO}}$ and $\Gamma$. Although a large deviation is present at lower 
frequencies, a statistically acceptable linear correlation is obtained, as the uncertainties on the fitted parameters 
remain well constrained. According to \citep{ST07,ST09}, these two parameters follow a scaling relation that has been 
used to estimate the masses of many unknown sources. In a follow-up study, we will investigate this scaling relation 
in detail and present the results elsewhere.

%\subsubsection{Anti-correlation of HR with QPO Frequency and Photon Index}
%
%The hardness ratio (HR) exhibits a trend opposite to that of the QPO frequency (see Fig.~\ref{figfig_lchr}). 
%To confirm this statistically, we analyzed the Pearson and Spearman rank correlations. Both methods indicate a strong 
%negative (or anti-) correlation between the QPO frequency and HR (see Fig.~\ref{fig_cfig_corr_qpohr}c). The Pearson 
%correlation coefficient is $P_{\text{cor}} = -0.653$ with a p-value of $5.1 \times 10^{-5}$, while the Spearman rank 
%correlation coefficient is $SR_{\text{cor}} = -0.738$ with a p-value of $1.5 \times 10^{-6}$.
%
%Similarly, while studying the correlation between HR and the photon index ($\Gamma$), as expected we also found a 
%strong anti-correlation between them (see Fig.~\ref{fig_cfig_corr_qpohr}d). The Pearson and Spearman rank correlation 
%coefficients are obtained as $P_{\text{cor}} = -0.629$ (p-value = $2.1 \times 10^{-7}$) and 
%$SR_{\text{cor}} = -0.627$ (p-value = $2.3 \times 10^{-7}$), respectively.

\begin{figure}[!t]
\vskip -0.3cm
  \centering
    \includegraphics[angle=0,width=9.0cm,keepaspectratio=true]{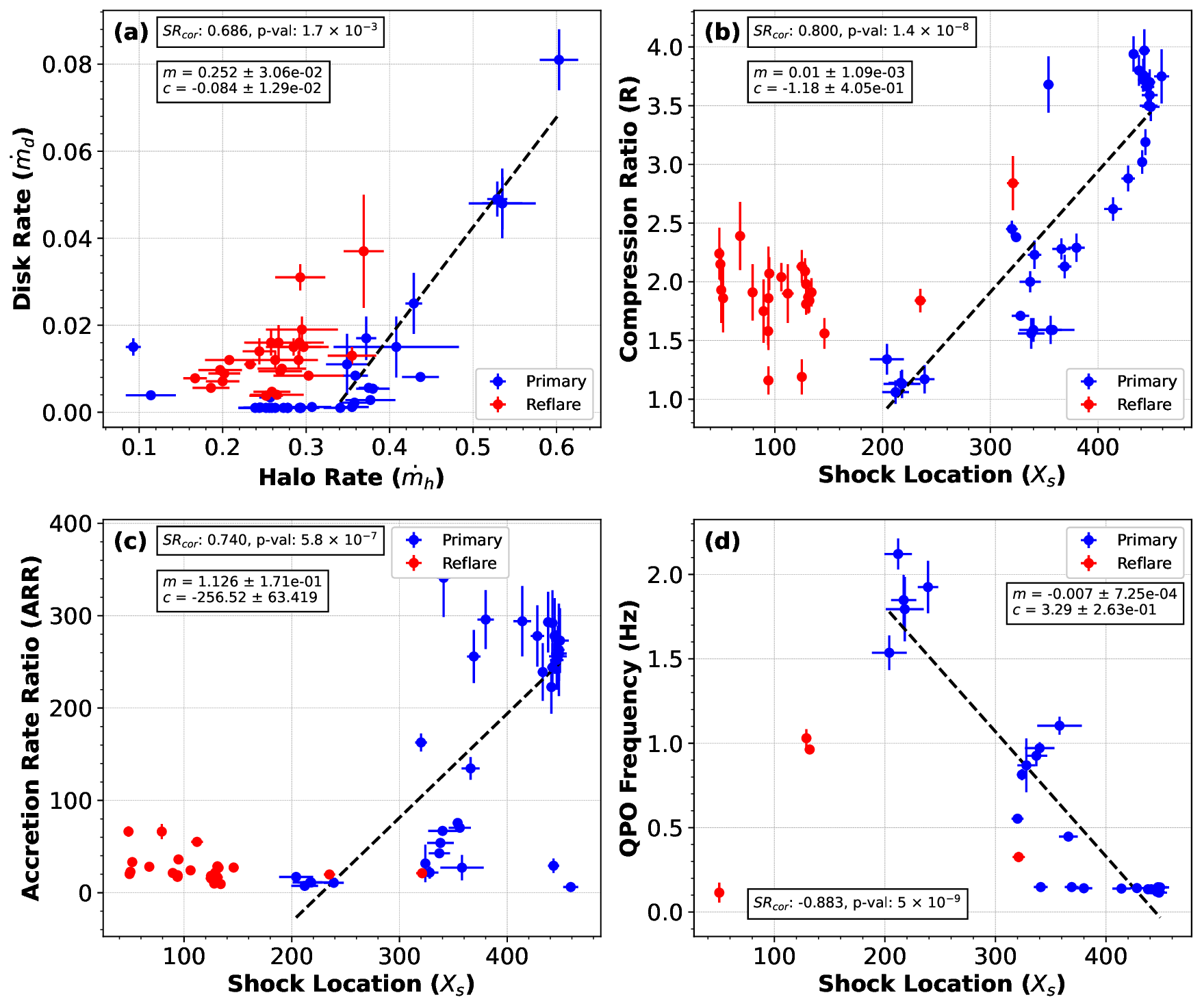}
         \caption{The variation of TCAF model-fitted (a) sub-Keplerian halo rate ($\dot{m}_h$ in $\dot{M}_{\rm Edd}$) 
         with Keplerian disk rate ($\dot{m}_d$ in $\dot{M}_{\rm Edd}$), (b) shock location ($X_s$ in $r_s$) with shock 
         compression ratio ($R$), shown in the top two plots, while in the bottom two plots, the variation of shock 
         location with the accretion rate ratio ($ARR = \dot{m}_h/\dot{m}_d$) and QPO frequency ($\nu_{\rm QPO}$) is 
         shown. The blue points represent the primary outburst (combining Phases I-III), and the red points correspond 
         to the rebrightening phase of the outburst (Phase IV). Spearman rank correlation coefficients 
         ($SR_{\rm cor}$), along with their $p$-values, are shown in the insets. The dashed black lines represent linear 
         ($y = mx + c$) trends for the evolution at higher halo accretion rates ($\dot{m}_h > 0.33~\dot{M}_{\rm Edd}$) 
         in panel (a) and at larger shock locations ($X_s > 200~r_s$) in the remaining panels. 
         }
\label{fig_corr_tcaf}
\end{figure}

\subsubsection{Correlation between TCAF Model Fitted Flow Parameters}

TCAF model fits directly provide the accretion flow parameters ($\dot{m}_d$, $\dot{m}_h$, $X_s$, $R$). In Fig.~\ref{fig_corr_tcaf}, 
the correlations among these parameters are examined. The blue points represent the model-fitted parameter values from the primary 
outburst (Phases I–III), while the red points correspond to the the rebrightening phase (Phase IV) of the outburst of 
MAXI~J1834$-$021. The correlations among the TCAF model-fitted flow parameters are complex and strongly dependent on the spectral 
state. Consequently, the flow parameters exhibit significant variations and differ between the primary and reflare episodes of 
the outburst.

For higher halo accretion rates ($\dot{m}_h > 0.33~\dot{M}_{\rm Edd}$), the disk and halo accretion rates exhibit a 
strong positive correlation. A broadly similar trend is observed for the evolution of shock location $X_s$ with the compression 
ratio $R$ and accretion rate ratio ARR at higher radii ($X_s > 200~r_s$). A linear fit to these conditional variations among 
the flow parameters is found to be statistically significant and is also shown in Fig.~\ref{fig_corr_tcaf}(a--c). Notably, 
the disk rate is found to lie within a relatively narrow range for lower halo rates ($<0.33~\dot{M}_{\rm Edd}$). A roughly 
similar trend in the evolution of $R$ and ARR is observed when the shock is formed closer ($<200~r_s$) to the BH. 
These characteristics of the TCAF flow parameters are observed predominantly during the reflare episode.

Furthermore, we find that the observed QPO frequency exhibits a strong negative correlation with the shock location, particularly 
for $X_s > 200~r_s$. A linear fit to this variation provides a reasonable approximation (Fig.~\ref{fig_corr_tcaf}d). However, according 
to SOM, the relationship between these two parameters is not linear, but instead follows $\nu_{\rm QPO} \sim X_s^{-3/2}$. 
Notably, a significant deviation is observed for three red points, where the shock locations fall below $200~r_s$ during the late 
phase of the outburst. This suggests that the origin of these QPOs may differ from that of the others.

A Spearman rank correlation analysis of the above-mentioned conditional ($\dot{m}_h > 0.33~\dot{M}_{\rm Edd}$ and $X_s > 200~r_s$) 
variations among the TCAF model-fitted accretion flow parameters and the QPO frequency shows trends consistent with those inferred from the 
linear fits. We find strong positive correlations for $\dot{m}_h$ vs.\ $\dot{m}_d$, $X_s$ vs.\ ARR, and $X_s$ vs.\ $R$, 
while a strong anti-correlation is observed between $X_s$ and $\nu_{\rm QPO}$.

Note that the observations primarily responsible for the strong correlations among the TCAF model-fitted parameters at 
higher halo accretion rates or larger shock radii belong to the primary episode of the outburst. This suggests that the accretion 
flow properties, and hence the source behavior, differed significantly between the primary and reflare episodes.

\section{Discussion and Concluding Remarks}

A detailed study of the spectral and timing properties of the newly discovered Galactic transient BHC MAXI~J1834$-$021 
is presented for the initial $\sim 8$ months (2023 March 7 to October 4; MJD = 60010.01--60221.37) of its discovery outburst. 
Archival data from NICER, obtained on a roughly daily basis in the $0.5$--$10$~keV band, are used in this study. Based on the 
evolution of the X-ray intensity, hardness ratio, QPO frequency, and spectral model-fitted parameters, we classify the entire 2023 
outburst of MAXI~J1834$-$021 into four phases. Broadly, the multi-peaked outburst consists of two main episodes: a primary episode 
and a reflare (for more details, see rebrightening criteria of \citet{Zhang19}). The Phases I--III constitute the primary episode, 
while Phase IV of the 2023 activity of MAXI~J1834$-$021 can be classified as a reflare (or rebrightening) phase of the outburst.

The outburst profiles of the primary episode in three energy bands—soft ($0.5$--$3$~keV, SXR), hard ($3$--$10$~keV, HXR), 
and total ($0.5$--$10$~keV, TXR)—along with the hardness ratio (defined as the ratio of HXR to SXR), indicate that, except for the 
initial two days, the source behaves like the decay phase of canonical transient BHCs. 
Due to the unavailability of NICER observations for $\sim 1$ month during the initial phase, it appears that an important 
onset phase of the outburst has been missed. The short duration of Phase I likely corresponds to this rising phase. 
The monotonically increasing trend of the QPO frequency in Phase~I and the decreasing trends in Phases~II and III are also 
consistent with the evolution of LFQPOs in the harder spectral states during the rising and declining phases of outbursts 
in transient BHCs \citep[][and references therein]{D13,Nandi12}. Due to the limited number of QPO detections in the reflare episode 
i.e., in Phase-IV, we are unable to comment on any systematic QPO trend.
%Furthermore, as reported in our earlier studies (see, e.g., \citep{D13,Nandi12}), such monotonic evolution of QPOs is typically 
%observed in the harder spectral states (LHS and HIMS) of an outburst, which is later confirmed by our spectral analysis. 
Notably, due to the faster decay of SXR (compared to HXR), an increasing trend in HR is observed in Phases II \& III, which moves 
the source toward a more harder spectral states. The evolution of the HID (Fig.~\ref{fig_hid}) during the primary episode further 
confirms that the source was in the declining phase of the outburst.

A detailed spectral analysis using both phenomenological (DBB+PL or PL) and physical (TCAF) models provides a clearer understanding 
of the accretion flow dynamics of the source during the 2023 outburst. 
%The absence of the thermal DBB component is observed during Phase III of the outburst. 
%The evolution of the spectral parameters ($T_{\rm in}$ and $\Gamma$), along with the model component 
%fluxes, provides a clearer understanding of the source behavior across different phases. 
The observed $\Gamma < 2.16$ throughout the outburst implies that the source remained in harder spectral states during the outburst. 
The evolution of $T_{\rm in}$, $\Gamma$, and the model-component fluxes, together with the temporal properties including 
the nature of the QPOs, enables us to understand the spectral behavior of the source across different phases of the outburst. 
Phase I is identified as the rising HIMS, Phase II as the declining HIMS, Phase III as the declining LHS, and Phase IV as an 
unclassified HIMS/LHS. The observation of two simultaneous HIMS phases in Phases I \& II is not a new phenomenon. 
\citet{Jana16} reported a similar case of two consecutive HIMS phases in the 2012–13 `failed' outburst of MAXI~J1836$-$194. 
Notably, during the outburst, the disk component remains thermally cool, as indicated by the relatively low values of $T_{\rm in}$.

%In our previous studies for many transient BHCs (for example during 2010-11 outburst of GX 339-4; 2010 \& 2011 outbursts 
%of H~1743-322; see \citep{D13,N12}), peak of the monotonically evolving QPOs is seen on source transits 
%HIMS $\leftrightarrow$ SIMS.
%The evolution of QPOs clearly signifies that the source remained in harder spectral states throughout the outburst. 
%Additionally, it indicates that the source was in HIMS (rising) during Phase-I, HIMS (declining) in Phase-II, and LHS 
%(declining) in Phase-III. These findings are further confirmed through spectral analysis. In a classical or type-I 
%outburst, the peak of the evolving QPO is generally observed on the HIMS-to-SIMS transition day \citep[for a review, 
%see][]{D18}. However, due to the absence of SIMS in this case, we observe the QPO peak on the transition day between 
%two HIMS. During the secondary outburst (i.e., in Phase-IV), based on the observation of four QPOs and their trends, 
%we can tentatively state that, similar to the primary outbursting phase, the QPO frequency initially increases before 
%decreasing in the later phase.

%%%%%%%%%%%%
%The TCAF model fits allow us to understand the nature of the source from a physical point of view. 
The evolution of the TCAF model-fitted/derived accretion flow parameters,
%, such as the Keplerian disk rate, sub-Keplerian halo rate, shock location, shock strength and ARR, 
allowed us to gain more insights into the accretion flow dynamics of the source during different phases of the 
outburst. The nature of the flow parameters strongly confirms the presence of double outbursts during the entire 2023 
epoch of MAXI~J1834-021. The high dominance of the sub-Keplerian halo accretion rate over the Keplerian disk rate throughout 
the outburst is consistent with the presence of the harder spectral states. In the declining Phase II, within $\sim 5$~days 
from the peak rate day, %(MJD=600012; the transition day of Phase I \& II, i.e., two HIMS) on March 14, 2023 (MJD=60017.23), 
the disk accretion rate is found to fall to its lower values, indicating the switching off of the viscous processes 
at the outer radius of the disk, i.e., at the pile-up radius $X_p$. Due to this low viscosity, the total accumulated 
matter at the pile-up radius prior to the outburst was not cleared in the primary episode of the source alone. 
The leftover matter might be fully or partially cleared in the reflare phase of the outburst.

A `kink' in the disk rate on May 18, 2023 (MJD=60082.68) indicates the possible triggering of the reflare 
outburst (Phase-IV). After that, we see the shock moving inward with a weakening strength, which is consistent with
the rising phase of an outburst of transient BHCs. The presence of the receding shock with increasing $R$ in 
Phases II \& III, where we observed a monotonically evolving QPO frequency, is consistent with the prescription 
mentioned in the SOM. According to the SOM, $\nu_{\rm QPO}$ is inversely proportional to 
the matter infall time scales and with the shock location as $X_s^{-3/2}$. 
Notably, on this `kink' observation, the PL model-fitted spectrum also indicates the softening of the spectra 
with a lower PL flux and a jump in $\Gamma$ value. 
Furthermore, we made preliminary estimates of the QPO frequencies for the initial few observations using the shock and mass parameters obtained from the TCAF model fits, following the prescription of \citet{D14}. The reasonable agreement between the estimated and observed values of $\nu_{\rm QPO}$ supports their origin in resonance shock oscillations. A detailed investigation of this aspect will be presented elsewhere.

The physical TCAF model fits also provide an estimate of the probable mass of the source since $M_{\rm BH}$ is a model 
parameter. The mass of the black hole is inherently linked to the model, as it influences the flow parameters such as 
the mass accretion rates, and the shock location and shock strength. From the model fits, we find a possible $M_{\rm BH}$ 
in the range of $11.9$--$13~M_\odot$, with an average value of $12.4~M_\odot$, and standard deviation of 
$1.7~M_\odot$. Thus we infer probable mass of the BH $M_{\rm BH} = 12.4\pm1.7~M_\odot$. This estimated $M_{\rm BH}$ 
is further supported by other independent methods \citep{DC26}.

The correlations among the spectral model-fitted parameters, SXR, and QPO frequencies provide further insight 
into the nature of the source. A strong positive correlation between SXR and $\nu_{\rm QPO}$ indicates that the thermal 
flux from the Keplerian disk plays an important role in the evolution of the QPO. An increasing Keplerian disk rate 
enhances the cooling of the hot corona, which in turn pushes the shock toward the BH by reducing the size of the corona 
(or CENBOL). As $X_s$ decreases with increasing SXR, a corresponding rise in $\nu_{\rm QPO}$ is observed. The TCAF 
model-fitted disk rate shows a strong positive correlation with higher halo rates ($\dot{m}_h > 0.33~\dot{M}_{\rm Edd}$). 
Similarly, the shock location exhibits strong positive correlations with the shock compression ratio and the 
accretion rate ratio, while showing a strong anti-correlation with the QPO frequency at larger radii.
This is consistent with the SOM framework, where $\nu_{\rm QPO} \sim X_s^{-3/2}$. During the rising phase, where 
a monotonically increasing trend in the QPO frequency is observed, enhanced cooling weakens 
the shock (reducing $R$) as it moves inward. Conversely, during the declining phase, where a monotonically decreasing trend 
in the QPO frequency is observed, the shock moves outward with a possible increase in strength.

According to the pile-up concept proposed by Chakrabarti and his collaborators, a large amount of matter was 
accumulated during the quiescence phase prior to the 2023 activity of the source. Since the primary outburst (up to 
Phase III) was a `failed' one, in which only harder spectral states were observed, the entire accumulated matter was 
not cleared in this primary episode. The remaining matter might have been partially or completely cleared 
during the secondary i.e., rebrightening episode i.e., in Phase-IV. A similar scenario was observed during the 
2004 outburst of H~1743-322 \citep{C19} and the 2013 or 2017–18 outburst of GX~339-4 \citep{Bhowmick21}, where the 
accumulated matter prior to these outbursts was not fully cleared and was instead expelled during the subsequent 
complete outburst.

Recently, \citet{Manca25} reported the spectral and timing properties of MAXI~J1834$-$021 mainly using the 
\texttt{nthComp} model, whereas in this {\it paper}, we present a more in-depth study to understand the accretion 
flow dynamics of the source during its discovery outburst. In contrast to their analysis, our spectro-temporal 
analysis of the source using both physical and phenomenological models not only allows us to infer the spectral nature 
of the source across four distinct observational phases from a physical perspective, but also provides insights into 
the possible origin of the reflare. We also estimate the mass of this newly discovered BHC within reasonable statistical 
limits from spectral fitting using the TCAF model.

\section*{Acknowledgements}

We are thankful to the anonymous referees for their kind suggestions to improve the quality of the paper.
This work made use of NICER/XTI data supplied by the High Energy Astrophysics Science Archive 
Research Center (HEASARC) archive. D.D. acknowledge the visiting research grant of National Tsing Hua University, 
Taiwan (NSTC 114-2811-M-007-084). H.-K. C. is supported by NSTC of Taiwan under grant NSTC 114-2112-M-007-042.

%\bibliographystyle{mnras}
%\bibliography{referencesBH} 

%%%%%%%%%%
\bibliographystyle{elsarticle-harv}
\bibliography{ddrefs}
%%%%%%%%%%

\begin{table*}%[!h]
\renewcommand{\thetable}{A\arabic{table}} % Makes numbering A1, A2, etc.
\setcounter{table}{0} % Reset counter so first appendix table is A1
%\addtolength{\tabcolsep}{-3.0pt} % to reduce gap between columns
\vskip -0.0cm
\centering
%\caption{Primary QPO Information}
\caption{Rate, Hardness and QPO Results}
\label{table_rate_qpo}
\small
%\scriptsize
\vskip -0.3cm
\renewcommand{\arraystretch}{0.9}  % to reduce gap between rows
\begin{tabular}{lcccc|cccc|cccc} 
%\begin{tabular}{|l|cc||c|c|c|c|c|} 
 \hline
	Sr.& ObsID & UT & MJD$_{Avg}$ & Exp & TXR & SXR & HXR & HR & $\nu_{QPO}$ & Q & rms(\%) & $\sigma$ \\
%No.&       & (mm/dd) &             & (s) & (Hz)        &   &          \\
\multicolumn{5}{c|}{\textsc{Data Information}}& \multicolumn{4}{c|}{\textsc{Rate and HR}}& \multicolumn{4}{c}{\textsc{QPO Result}} \\
\hline
 1&6203690101&03/07& 60010.01& 9545 & 44.9$^{\pm0.22}$&35.6$^{\pm0.19}$&9.47$^{\pm0.20}$&0.27$^{\pm0.01}$&1.848$^{\pm0.149}$&3.69&18.6&2.17 \\ 
 2&6203690102&03/08& 60011.03& 6053 & 43.0$^{\pm0.27}$&35.8$^{\pm0.24}$&7.23$^{\pm0.24}$&0.20$^{\pm0.01}$&1.794$^{\pm0.191}$&4.11&11.2&2.12 \\
 3&6203690103&03/08& 60011.99& 7471 & 46.9$^{\pm0.25}$&37.9$^{\pm0.22}$&9.11$^{\pm0.24}$&0.24$^{\pm0.01}$&2.121$^{\pm0.092}$&3.79&11.0&3.28 \\
 4&6203690104&03/10& 60013.23& 7602 & 43.7$^{\pm0.24}$&34.7$^{\pm0.21}$&9.05$^{\pm0.23}$&0.26$^{\pm0.01}$&1.925$^{\pm0.155}$&2.53&12.2&2.99 \\
 5&6203690105&03/11& 60014.13& 3895 & 35.6$^{\pm0.30}$&29.3$^{\pm0.27}$&6.32$^{\pm0.28}$&0.22$^{\pm0.01}$&1.536$^{\pm0.102}$&2.84&12.9&3.54 \\
 6&6203690106&03/12& 60015.36& 1274 & 40.4$^{\pm0.63}$&32.8$^{\pm0.51}$&7.74$^{\pm0.76}$&0.23$^{\pm0.02}$& ---              & ---& ---&---  \\
 7&6203690107&03/13& 60016.00& 5429 & 27.1$^{\pm0.22}$&22.1$^{\pm0.20}$&4.98$^{\pm0.21}$&0.22$^{\pm0.01}$&1.104$^{\pm0.053}$&1.64&18.9&6.36 \\
 8&6203690108&03/14& 60017.23& 3525 & 24.1$^{\pm0.26}$&19.5$^{\pm0.24}$&4.62$^{\pm0.25}$&0.24$^{\pm0.01}$&0.971$^{\pm0.036}$&2.97&15.5&4.02 \\
 9&6203690109&03/15& 60018.26& 3505 & 30.1$^{\pm0.32}$&24.4$^{\pm0.26}$&5.80$^{\pm0.37}$&0.24$^{\pm0.02}$& ---              &--- &--- &---  \\  
10&6203690110&03/16& 60019.24& 2547 & 29.5$^{\pm0.34}$&24.1$^{\pm0.31}$&5.47$^{\pm0.31}$&0.23$^{\pm0.01}$& ---              &--- &--- &---  \\ 
11&6203690111&03/17& 60020.00& 2969 & 25.2$^{\pm0.29}$&20.5$^{\pm0.26}$&4.70$^{\pm0.29}$&0.23$^{\pm0.01}$& ---              &--- &--- &---  \\ 
12&6203690112&03/18& 60021.64& 2165 & 22.5$^{\pm0.32}$&17.8$^{\pm0.29}$&4.70$^{\pm0.32}$&0.26$^{\pm0.02}$&0.926$^{\pm0.062}$&1.96&18.2&3.91 \\ 
13&6203690113&03/19& 60022.15& 1712 & 21.2$^{\pm0.35}$&17.0$^{\pm0.31}$&4.21$^{\pm0.35}$&0.25$^{\pm0.02}$&0.869$^{\pm0.159}$&1.28&20.4&3.43 \\
14&6203690114&03/20& 60023.36& 3294 & 19.9$^{\pm0.25}$&16.0$^{\pm0.22}$&3.91$^{\pm0.24}$&0.24$^{\pm0.02}$&0.814$^{\pm0.033}$&5.25&12.5&2.75 \\
15&6203690115&03/21& 60024.07& 4974 & 19.2$^{\pm0.20}$&14.5$^{\pm0.17}$&4.78$^{\pm0.20}$&0.33$^{\pm0.01}$&0.749$^{\pm0.049}$&1.72&19.4&5.07 \\
16&6203690116&03/22& 60025.43& 3255 & 15.2$^{\pm0.22}$&11.9$^{\pm0.19}$&3.27$^{\pm0.20}$&0.27$^{\pm0.02}$&0.553$^{\pm0.020}$&3.19&18.5&4.01 \\
17&6203690117&03/23& 60026.73& 1715 & 14.3$^{\pm0.38}$&10.7$^{\pm0.33}$&3.62$^{\pm0.40}$&0.34$^{\pm0.04}$&0.464$^{\pm0.018}$&7.88&14.0&2.24 \\
18&6203690118&03/24& 60027.11& 2557 & 17.7$^{\pm0.31}$&13.5$^{\pm0.23}$&4.24$^{\pm0.43}$&0.31$^{\pm0.03}$&0.447$^{\pm0.027}$&4.03&12.5&2.39 \\
19&6203690119&03/25& 60028.78& 1383 & 12.9$^{\pm0.31}$&9.00$^{\pm0.25}$&4.00$^{\pm0.33}$&0.44$^{\pm0.04}$& ---              & ---& ---&---  \\
20&6203690120&03/26& 60029.30& 1582 & 10.6$^{\pm0.26}$&8.15$^{\pm0.23}$&2.51$^{\pm0.26}$&0.31$^{\pm0.03}$& ---              & ---& ---&---  \\ 
21&6203690122&03/28& 60031.45& 1051 & 9.74$^{\pm0.30}$&7.55$^{\pm0.27}$&2.21$^{\pm0.32}$&0.29$^{\pm0.04}$& ---              & ---& ---&---  \\ 
22&6203690125&04/05& 60039.57& 2692 & 7.96$^{\pm0.17}$&5.19$^{\pm0.14}$&2.78$^{\pm0.20}$&0.54$^{\pm0.04}$&0.175$^{\pm0.008}$&3.27&22.6&3.37 \\ 
23&6203690126&04/09& 60043.26& 3353 & 6.98$^{\pm0.14}$&4.42$^{\pm0.11}$&2.58$^{\pm0.17}$&0.58$^{\pm0.04}$&0.148$^{\pm0.007}$&3.70&21.7&3.23 \\ 
24&6203690127&04/10& 60044.15& 1540 & 6.55$^{\pm0.21}$&4.31$^{\pm0.17}$&2.26$^{\pm0.24}$&0.52$^{\pm0.06}$&0.148$^{\pm0.008}$&5.21&20.2&2.01 \\ 
25&6203690128&04/11& 60045.18& 1133 & 6.69$^{\pm0.24}$&3.36$^{\pm0.17}$&3.34$^{\pm0.34}$&0.99$^{\pm0.11}$&0.141$^{\pm0.010}$&6.46&19.7&1.76 \\ 
26&6203690129&04/12& 60046.67& 2609 & 9.50$^{\pm0.19}$&5.79$^{\pm0.15}$&3.72$^{\pm0.23}$&0.64$^{\pm0.04}$& ---              & ---& ---&---  \\ 
27&6203690130&04/13& 60047.52& 1826 & 4.29$^{\pm0.15}$&3.12$^{\pm0.13}$&1.18$^{\pm0.17}$&0.38$^{\pm0.06}$&0.139$^{\pm0.007}$&4.17&21.6&1.98 \\      
28&6203690131&04/14& 60048.10& 2292 & 4.57$^{\pm0.14}$&3.35$^{\pm0.12}$&1.24$^{\pm0.15}$&0.37$^{\pm0.05}$&0.142$^{\pm0.006}$&5.73&21.2&2.41 \\      
29&6203690132&04/15& 60049.20& 1005 & 4.59$^{\pm0.21}$&3.36$^{\pm0.18}$&1.23$^{\pm0.23}$&0.37$^{\pm0.07}$&0.135$^{\pm0.010}$&5.04&21.3&1.75 \\      
30&6203690133&04/17& 60051.25& 1058 & 5.10$^{\pm0.22}$&3.63$^{\pm0.18}$&1.48$^{\pm0.24}$&0.41$^{\pm0.07}$& ---              & ---& ---&---  \\
31&6203690134&04/18& 60052.67& 1847 & 14.1$^{\pm0.28}$&7.89$^{\pm0.21}$&6.29$^{\pm0.37}$&0.80$^{\pm0.05}$&0.147$^{\pm0.007}$&6.10&17.9&2.21 \\      
32&6203690135&04/20& 60054.54& 3140 & 4.19$^{\pm0.11}$&2.96$^{\pm0.10}$&1.24$^{\pm0.14}$&0.42$^{\pm0.05}$&0.146$^{\pm0.014}$&2.48&18.7&2.51 \\      
33&6203690136&04/21& 60055.31& 2207 & 4.28$^{\pm0.14}$&3.09$^{\pm0.12}$&1.20$^{\pm0.15}$&0.39$^{\pm0.05}$&0.147$^{\pm0.014}$&3.02&18.7&2.76 \\
34&6203690137&04/26& 60060.22& 3109 & 4.20$^{\pm0.12}$&2.97$^{\pm0.10}$&1.23$^{\pm0.13}$&0.41$^{\pm0.05}$&0.115$^{\pm0.005}$&3.51&20.2&3.09 \\
35&6203690138&04/27& 60061.32& 1567 & 4.99$^{\pm0.18}$&3.29$^{\pm0.14}$&1.70$^{\pm0.21}$&0.52$^{\pm0.07}$&0.121$^{\pm0.015}$&2.70&19.0&1.98 \\
36&6203690139&04/28& 60062.29& 795  & 5.19$^{\pm0.25}$&3.52$^{\pm0.21}$&1.68$^{\pm0.29}$&0.48$^{\pm0.09}$&0.119$^{\pm0.009}$&5.34&17.6&1.61 \\
37&6203690140&04/29& 60063.58& 700  & 5.10$^{\pm0.27}$&3.45$^{\pm0.22}$&1.66$^{\pm0.32}$&0.48$^{\pm0.10}$& ---              & ---& ---&---  \\ 
38&6203690141&05/01& 60065.38& 710  & 4.35$^{\pm0.30}$&3.10$^{\pm0.25}$&1.26$^{\pm0.35}$&0.41$^{\pm0.12}$& ---              & ---& ---&---  \\ 
39&6203690142&05/06& 60070.48& 1679 & 8.44$^{\pm0.22}$&4.40$^{\pm0.16}$&4.05$^{\pm0.31}$&0.92$^{\pm0.08}$& ---              & ---& ---&---  \\ 
40&6203690143&05/12& 60076.57& 729  & 3.85$^{\pm0.23}$&2.74$^{\pm0.19}$&1.12$^{\pm0.26}$&0.41$^{\pm0.10}$& ---              & ---& ---&---  \\ 
41&6203690144&05/14& 60078.35& 349  & 11.6$^{\pm0.58}$&4.74$^{\pm0.37}$&6.87$^{\pm0.94}$&1.45$^{\pm0.23}$&0.135$^{\pm0.005}$&6.89&18.0&1.74 \\
42&6203690145&05/15& 60079.01& 1933 & 3.57$^{\pm0.14}$&2.50$^{\pm0.11}$&1.08$^{\pm0.16}$&0.43$^{\pm0.07}$& ---              & ---& ---&---  \\ 
43&6203690146&05/16& 60080.05& 809  & 3.53$^{\pm0.21}$&2.49$^{\pm0.18}$&1.04$^{\pm0.23}$&0.42$^{\pm0.10}$& ---              & ---& ---&---  \\ 
44&6203690147&05/17& 60081.84& 615  & 3.01$^{\pm0.22}$&2.17$^{\pm0.19}$&0.85$^{\pm0.26}$&0.39$^{\pm0.13}$& ---              & ---& ---&---  \\ 
45&6203690148&05/18& 60082.68& 866  & 2.52$^{\pm0.17}$&1.88$^{\pm0.15}$&0.64$^{\pm0.19}$&0.34$^{\pm0.10}$& ---              & ---& ---&---  \\ 
46&6203690149&05/19& 60083.26& 713  & 2.42$^{\pm0.18}$&1.78$^{\pm0.16}$&0.64$^{\pm0.19}$&0.36$^{\pm0.11}$& ---              & ---& ---&---  \\ 
47&6203690150&05/20& 60084.23& 890  & 5.30$^{\pm0.24}$&2.93$^{\pm0.18}$&2.38$^{\pm0.33}$&0.81$^{\pm0.12}$& ---              & ---& ---&---  \\ 
48&6203690151&05/21& 60085.07& 1213 & 2.51$^{\pm0.14}$&1.41$^{\pm0.11}$&1.10$^{\pm0.20}$&0.78$^{\pm0.15}$& ---              & ---& ---&---  \\ 
49&6203690152&05/22& 60086.43& 1256 & 2.83$^{\pm0.15}$&1.65$^{\pm0.11}$&1.18$^{\pm0.20}$&0.72$^{\pm0.13}$& ---              & ---& ---&---  \\ 
50&6203690153&05/23& 60087.46& 1941 & 2.28$^{\pm0.11}$&1.43$^{\pm0.09}$&0.86$^{\pm0.14}$&0.60$^{\pm0.11}$& ---              & ---& ---&---  \\ 

\hline
\end{tabular}
%\noindent{
\leftline{UT in mm/dd format. The units of Exposure (Exp) in s; rates (TXR, SXR, HXR) in cnts/s; $\nu_{QPO}$ in Hz. Q=$\nu_{QPO}$/$FWHM$, Sigma ($\sigma$) confidence.} 
% total ($0.5-10$~keV), soft ($0.5-3$~keV), hard ($3-10$~keV)
%\leftline{X (=6203690) is the initial part of ObsID; Q=$FWHM/\nu_{QPO}$; UT in mm/dd format. The units of Exposure (Exp) in s; }
%\leftline{$\nu_{QPO}$ in Hz; $N_{\rm H}$ in $10^{22}~{\rm cm}^{-2}$; $T_{in}$ in keV; model fluxes diskbb ($F_{\rm DBB}$) and pexrav ($F_{\rm pexr}$) in $10^{-8}~{\rm erg}~{\rm cm}^{-2}~{\rm s}^{-1}$.}
\end{table*}

\addtocounter{table}{-1} % <-- THIS keeps the same table number
\begin{table*}
\centering
\caption{Table A1 (continued): Rate, Hardness and QPO Results}
\small
\begin{tabular}{lcccc|cccc|cccc} 
\hline
	Sr.& ObsID & UT & MJD$_{Avg}$ & Exp & TXR & SXR & HXR & HR & $\nu_{QPO}$ & Q & rms(\%) & $\sigma$ \\
\multicolumn{5}{c|}{\textsc{Data Information}}& \multicolumn{4}{c|}{\textsc{Rate and HR}}& \multicolumn{4}{c}{\textsc{QPO Result}} \\
\hline
51&6203690154&05/29& 60093.39& 2717 & 1.59$^{\pm0.08}$&0.91$^{\pm0.06}$&0.67$^{\pm0.10}$&0.74$^{\pm0.12}$& ---              & ---& ---&---  \\ 
52&6203690155&06/08& 60103.97& 1020 & 8.30$^{\pm0.29}$&6.33$^{\pm0.25}$&1.98$^{\pm0.28}$&0.31$^{\pm0.05}$& ---              & ---& ---&---  \\ 
53&6203690157&06/14& 60109.38& 1790 & 15.3$^{\pm0.29}$&12.5$^{\pm0.26}$&2.82$^{\pm0.29}$&0.23$^{\pm0.02}$& ---              & ---& ---&---  \\ 
54&6203690158&06/17& 60112.03& 722  & 18.9$^{\pm0.51}$&14.7$^{\pm0.45}$&4.27$^{\pm0.50}$&0.29$^{\pm0.04}$&0.326$^{\pm0.030}$&6.72&7.40&1.74 \\
55&6203690159&06/18& 60113.26& 1112 & 18.4$^{\pm0.41}$&14.9$^{\pm0.37}$&3.53$^{\pm0.39}$&0.24$^{\pm0.03}$& ---              & ---& ---&---  \\ 
56&6203690160&06/19& 60114.23& 537  & 16.8$^{\pm0.56}$&13.6$^{\pm0.50}$&3.26$^{\pm0.53}$&0.24$^{\pm0.04}$& ---              & ---& ---&---  \\ 
57&6203690161&06/21& 60116.22& 1284 & 24.4$^{\pm0.44}$&16.2$^{\pm0.35}$&8.29$^{\pm0.50}$&0.51$^{\pm0.03}$& ---              & ---& ---&---  \\ 
58&6203690162&06/22& 60117.08& 854  & 16.8$^{\pm0.43}$&13.7$^{\pm0.39}$&3.06$^{\pm0.42}$&0.22$^{\pm0.03}$& ---              & ---& ---&---  \\ 
59&6203690163&06/23& 60118.09& 2803 & 17.5$^{\pm0.25}$&14.0$^{\pm0.22}$&3.49$^{\pm0.23}$&0.25$^{\pm0.02}$& ---              & ---& ---&---  \\ 
60&6203690164&06/24& 60119.20& 1383 & 27.1$^{\pm0.44}$&17.1$^{\pm0.35}$&10.0$^{\pm0.54}$&0.59$^{\pm0.03}$& ---              & ---& ---&---  \\ 
61&6203690165&06/25& 60120.16& 1755 & 26.3$^{\pm0.39}$&17.1$^{\pm0.31}$&9.31$^{\pm0.45}$&0.55$^{\pm0.03}$& ---              & ---& ---&---  \\ 
62&6203690166&06/27& 60122.36& 1632 & 18.0$^{\pm0.33}$&14.7$^{\pm0.30}$&3.32$^{\pm0.33}$&0.23$^{\pm0.02}$& ---              & ---& ---&---  \\ 
63&6203690167&06/29& 60124.29& 886  & 18.2$^{\pm0.45}$&14.4$^{\pm0.40}$&3.76$^{\pm0.44}$&0.26$^{\pm0.03}$& ---              & ---& ---&---  \\ 
64&6203690168&07/01& 60126.05& 1604 & 15.2$^{\pm0.31}$&12.3$^{\pm0.28}$&2.92$^{\pm0.29}$&0.24$^{\pm0.02}$& ---              & ---& ---&---  \\ 
65&6203690169&07/02& 60127.19& 1447 & 14.9$^{\pm0.32}$&12.1$^{\pm0.29}$&2.83$^{\pm0.30}$&0.23$^{\pm0.03}$&0.963$^{\pm0.012}$&3.72&12.9&1.84 \\
66&6203690170&07/03& 60128.42& 1462 & 15.8$^{\pm0.33}$&13.0$^{\pm0.30}$&2.83$^{\pm0.30}$&0.22$^{\pm0.02}$& ---              & ---& ---&---  \\ 
67&6203690171&07/04& 60129.20& 1763 & 14.2$^{\pm0.28}$&11.4$^{\pm0.25}$&2.74$^{\pm0.28}$&0.24$^{\pm0.02}$&1.030$^{\pm0.053}$&4.84&15.0&2.36 \\
68&6203690172&07/06& 60131.98& 551  & 14.9$^{\pm0.52}$&11.8$^{\pm0.46}$&3.09$^{\pm0.51}$&0.26$^{\pm0.04}$& ---              & ---& ---&---  \\ 
69&6203690173&07/09& 60134.31& 2046 & 12.5$^{\pm0.25}$&9.96$^{\pm0.22}$&2.60$^{\pm0.24}$&0.26$^{\pm0.02}$& ---              & ---& ---&---  \\ 
70&6203690179&07/16& 60141.01& 1985 & 13.1$^{\pm0.25}$&10.3$^{\pm0.23}$&2.77$^{\pm0.25}$&0.27$^{\pm0.03}$& ---              & ---& ---&---  \\ 
71&6203690180&07/17& 60142.43& 894  & 14.9$^{\pm0.41}$&10.7$^{\pm0.35}$&4.19$^{\pm0.45}$&0.39$^{\pm0.04}$& ---              & ---& ---&---  \\ 
72&6203690181&07/18& 60143.27& 784  & 14.9$^{\pm0.43}$&11.3$^{\pm0.38}$&3.62$^{\pm0.43}$&0.32$^{\pm0.04}$& ---              & ---& ---&---  \\ 
73&6203690185&07/22& 60147.66& 1882 & 12.4$^{\pm0.26}$&9.74$^{\pm0.23}$&2.65$^{\pm0.25}$&0.27$^{\pm0.03}$& ---              & ---& ---&---  \\ 
74&6203690186&07/23& 60148.24& 2247 & 13.0$^{\pm0.24}$&10.4$^{\pm0.22}$&2.62$^{\pm0.24}$&0.25$^{\pm0.02}$& ---              & ---& ---&---  \\ 
75&6203690187&07/24& 60149.01& 1732 & 12.2$^{\pm0.27}$&9.58$^{\pm0.24}$&2.61$^{\pm0.28}$&0.27$^{\pm0.03}$& ---              & ---& ---&---  \\ 
76&6203690192&07/29& 60154.51& 2579 & 19.2$^{\pm0.27}$&15.7$^{\pm0.25}$&3.51$^{\pm0.26}$&0.22$^{\pm0.02}$& ---              & ---& ---&---  \\ 
77&6203690193&07/30& 60155.40& 2545 & 19.7$^{\pm0.28}$&16.2$^{\pm0.25}$&3.49$^{\pm0.25}$&0.22$^{\pm0.02}$& ---              & ---& ---&---  \\ 
78&6203690195&08/01& 60157.01& 1689 & 21.9$^{\pm0.36}$&18.2$^{\pm0.33}$&3.76$^{\pm0.34}$&0.21$^{\pm0.02}$& ---              & ---& ---&---  \\ 
79&6203690196&08/02& 60158.88& 921  & 21.5$^{\pm0.48}$&17.1$^{\pm0.43}$&4.43$^{\pm0.48}$&0.26$^{\pm0.03}$& ---              & ---& ---&---  \\ 
80&6203690203&08/17& 60173.41& 2264 & 19.1$^{\pm0.29}$&13.1$^{\pm0.24}$&6.07$^{\pm0.33}$&0.46$^{\pm0.03}$& ---              & ---& ---&---  \\ 
81&6203690205&08/19& 60175.28& 2030 & 18.6$^{\pm0.30}$&12.4$^{\pm0.25}$&6.18$^{\pm0.35}$&0.50$^{\pm0.03}$& ---              & ---& ---&---  \\ 
82&6203690207&08/21& 60177.16& 188  & 13.6$^{\pm0.85}$&11.1$^{\pm0.76}$&2.54$^{\pm0.80}$&0.23$^{\pm0.07}$& ---              & ---& ---&---  \\ 
83&6203690209&08/26& 60182.97& 623  & 12.7$^{\pm0.45}$&10.5$^{\pm0.41}$&2.24$^{\pm0.42}$&0.21$^{\pm0.04}$& ---              & ---& ---&---  \\ 
84&6203690210&08/27& 60183.55& 536  & 12.3$^{\pm0.48}$&10.1$^{\pm0.43}$&2.31$^{\pm0.45}$&0.23$^{\pm0.05}$& ---              & ---& ---&---  \\ 
85&6203690213&08/30& 60186.18& 976  & 11.1$^{\pm0.34}$&8.99$^{\pm0.30}$&2.16$^{\pm0.34}$&0.24$^{\pm0.04}$& ---              & ---& ---&---  \\ 
86&6203690215&09/01& 60188.12& 5074 & 15.9$^{\pm0.18}$&11.2$^{\pm0.15}$&4.79$^{\pm0.20}$&0.43$^{\pm0.02}$&0.115$^{\pm0.059}$&5.84&11.9&1.85 \\ 
87&6203690216&09/02& 60189.02& 1232 & 20.7$^{\pm0.49}$&15.9$^{\pm0.35}$&4.83$^{\pm0.71}$&0.23$^{\pm0.05}$& ---              & ---& ---&---  \\ 
88&6203690218&09/04& 60191.02& 3076 & 13.2$^{\pm0.21}$&10.6$^{\pm0.18}$&2.60$^{\pm0.18}$&0.24$^{\pm0.02}$& ---              & ---& ---&---  \\ 
89&6203690219&09/05& 60192.51& 1137 & 12.5$^{\pm0.33}$&10.2$^{\pm0.29}$&2.37$^{\pm0.31}$&0.23$^{\pm0.03}$& ---              & ---& ---&---  \\ 
90&6203690224&09/10& 60197.74& 982  & 10.9$^{\pm0.33}$&8.93$^{\pm0.30}$&2.05$^{\pm0.32}$&0.23$^{\pm0.04}$& ---              & ---& ---&---  \\ 
91&6203690225&09/11& 60198.97& 547  & 10.4$^{\pm0.43}$&8.46$^{\pm0.39}$&1.93$^{\pm0.42}$&0.23$^{\pm0.05}$& ---              & ---& ---&---  \\ 
92&6203690226&09/12& 60199.74& 610  & 17.7$^{\pm0.53}$&10.6$^{\pm0.41}$&7.10$^{\pm0.68}$&0.67$^{\pm0.07}$& ---              & ---& ---&---  \\ 
93&6203690227&09/21& 60208.66& 457  & 8.78$^{\pm0.44}$&6.59$^{\pm0.38}$&2.21$^{\pm0.46}$&0.34$^{\pm0.07}$& ---              & ---& ---&---  \\ 
94&6203690228&09/24& 60211.57& 720  & 6.83$^{\pm0.31}$&5.26$^{\pm0.27}$&1.59$^{\pm0.32}$&0.30$^{\pm0.06}$& ---              & ---& ---&---  \\ 
95&6203690229&09/29& 60216.01& 1532 & 9.32$^{\pm0.25}$&7.49$^{\pm0.22}$&1.85$^{\pm0.24}$&0.25$^{\pm0.03}$& ---              & ---& ---&---  \\ 
96&6203690230&09/30& 60217.63& 974  & 10.5$^{\pm0.33}$&8.39$^{\pm0.29}$&2.22$^{\pm0.31}$&0.26$^{\pm0.04}$& ---              & ---& ---&---  \\ 
97&6203690231&10/02& 60219.44& 1292 & 8.79$^{\pm0.26}$&6.98$^{\pm0.23}$&1.82$^{\pm0.26}$&0.26$^{\pm0.04}$& ---              & ---& ---&---  \\ 
98&6203690232&10/04& 60221.37& 494  & 7.73$^{\pm0.40}$&6.12$^{\pm0.35}$&1.64$^{\pm0.39}$&0.27$^{\pm0.07}$& ---              & ---& ---&---  \\ 
\hline
\end{tabular}
\end{table*}

\begin{table*}%[!h]
\renewcommand{\thetable}{A\arabic{table}} % Makes numbering A1, A2, etc.
\addtolength{\tabcolsep}{-3.5pt} % to reduce gap between columns
\vskip -0.0cm
\centering
%\caption{Primary QPO Information}
\caption{Spectral Fitted Results}
\label{table_spec}
\small
%\scriptsize
\vskip -0.3cm
\renewcommand{\arraystretch}{0.9}  % to reduce gap between rows
\begin{tabular}{lcccccc|ccccccc} 
%\begin{tabular}{|l|cc||c|c|c|c|c|} 
 \hline
\multicolumn{7}{c|}{\textsc{DISKBB+PL}}& \multicolumn{7}{c}{\textsc{TCAF}} \\
Sr.& ObsID & T$_{\rm in}$ & $\Gamma$ & $\chi^2$/DOF & F$_{\rm DBB}$ & F$_{\rm PL}$ &  $\dot{m}_d$ & $\dot{m}_h$ & M$_{\rm BH}$ & $X_s$ & $R$ & N$_{\rm TCAF}$&$\chi^2$/DOF \\
\hline
 1&X101&0.209$^{\pm0.005}$&2.126$^{\pm0.091}$& 85/126&1.70$^{\pm0.02}$&2.47$^{\pm0.012}$& 0.048$^{\pm 0.008}$&0.535$^{\pm0.040}$&12.10$^{\pm0.10}$&217$^{\pm11.4}$&1.14$^{\pm0.10}$&3.88$^{\pm0.04}$& 86/123 \\
 2&X102&0.216$^{\pm0.004}$&2.060$^{\pm0.019}$&123/121&1.85$^{\pm0.02}$&2.48$^{\pm0.010}$& 0.048$^{\pm 0.006}$&0.535$^{\pm0.024}$&12.06$^{\pm0.12}$&218$^{\pm17.1}$&1.13$^{\pm0.12}$&4.09$^{\pm0.04}$&145/123 \\
 3&X103&0.223$^{\pm0.003}$&2.021$^{\pm0.023}$&125/127&1.87$^{\pm0.02}$&2.55$^{\pm0.010}$& 0.081$^{\pm 0.007}$&0.603$^{\pm0.023}$&12.97$^{\pm0.10}$&212$^{\pm12.4}$&1.06$^{\pm0.10}$&1.96$^{\pm0.04}$&146/123 \\
 4&X104&0.203$^{\pm0.004}$&2.158$^{\pm0.026}$& 88/128&1.58$^{\pm0.02}$&2.44$^{\pm0.011}$& 0.049$^{\pm 0.004}$&0.529$^{\pm0.012}$&12.12$^{\pm0.12}$&239$^{\pm 9.2}$&1.17$^{\pm0.12}$&4.33$^{\pm0.08}$& 83/125 \\
 5&X105&0.204$^{\pm0.004}$&2.027$^{\pm0.020}$&125/117&1.50$^{\pm0.02}$&2.12$^{\pm0.010}$& 0.025$^{\pm 0.007}$&0.429$^{\pm0.010}$&12.03$^{\pm0.15}$&204$^{\pm15.7}$&1.34$^{\pm0.13}$&6.35$^{\pm0.06}$&121/114 \\
 7&X107&0.203$^{\pm0.003}$&1.905$^{\pm0.033}$&138/116&1.26$^{\pm0.01}$&1.65$^{\pm0.007}$& 0.015$^{\pm 0.007}$&0.408$^{\pm0.075}$&12.99$^{\pm0.13}$&358$^{\pm20.1}$&1.59$^{\pm0.03}$&9.44$^{\pm0.09}$&141/117 \\
 8&X108&0.205$^{\pm0.004}$&1.845$^{\pm0.022}$&135/113&1.16$^{\pm0.02}$&1.47$^{\pm0.008}$&5.6E-3$^{\pm1.6E-4}$&0.375$^{\pm0.012}$&12.60$^{\pm0.12}$&340$^{\pm13.3}$&1.59$^{\pm0.10}$&28.4$^{\pm0.04}$&137/108 \\
 9&X110&0.203$^{\pm0.005}$&1.929$^{\pm0.025}$&129/112&1.34$^{\pm0.02}$&1.72$^{\pm0.010}$&8.1E-3$^{\pm1.0E-4}$&0.437$^{\pm0.022}$&12.64$^{\pm0.10}$&338$^{\pm12.3}$&1.56$^{\pm0.13}$&14.2$^{\pm0.04}$&140/109 \\
11&X111&0.196$^{\pm0.005}$&1.949$^{\pm0.023}$&121/114&1.10$^{\pm0.02}$&1.54$^{\pm0.009}$&5.4E-3$^{\pm1.0E-4}$&0.380$^{\pm0.020}$&12.96$^{\pm0.13}$&356$^{\pm10.3}$&1.59$^{\pm0.12}$&26.6$^{\pm0.05}$&107/110  \\
12&X112&0.186$^{\pm0.009}$&1.885$^{\pm0.053}$&128/112&0.71$^{\pm0.02}$&1.34$^{\pm0.009}$&8.4E-3$^{\pm1.0E-4}$&0.359$^{\pm0.015}$&12.80$^{\pm0.16}$&337$^{\pm9.82}$&2.00$^{\pm0.09}$&8.84$^{\pm0.02}$&126/110  \\
13&X113&0.184$^{\pm0.007}$&1.978$^{\pm0.031}$&118/107&0.89$^{\pm0.02}$&1.32$^{\pm0.010}$& 0.017$^{\pm 0.005}$&0.372$^{\pm0.012}$&12.98$^{\pm0.14}$&328$^{\pm8.05}$&1.71$^{\pm0.04}$&5.73$^{\pm0.10}$&116/104  \\
14&X114&0.202$^{\pm0.011}$&1.879$^{\pm0.024}$&137/114&0.75$^{\pm0.01}$&1.24$^{\pm0.007}$& 0.011$^{\pm 0.007}$&0.349$^{\pm0.025}$&12.38$^{\pm0.15}$&324$^{\pm4.47}$&2.38$^{\pm0.03}$&6.06$^{\pm0.01}$&129/110  \\
16&X116&0.205$^{\pm0.023}$&1.754$^{\pm0.026}$&111/114&0.54$^{\pm0.01}$&0.96$^{\pm0.006}$&2.2E-3$^{\pm0.7E-4}$&0.358$^{\pm0.018}$&12.20$^{\pm0.20}$&320$^{\pm5.35}$&2.45$^{\pm0.07}$&8.56$^{\pm0.06}$&116/111  \\
20&X120&0.213$^{\pm0.017}$&1.713$^{\pm0.046}$&116/106&0.27$^{\pm0.01}$&0.68$^{\pm0.007}$&2.8E-3$^{\pm1.2E-4}$&0.377$^{\pm0.030}$&12.50$^{\pm0.32}$&366$^{\pm8.40}$&2.28$^{\pm0.09}$&3.85$^{\pm0.08}$&109/102  \\
23&X126& ---              &1.679$^{\pm0.029}$&124/116&---             &0.26$^{\pm0.002}$&1.0E-3$^{\pm1.1E-4}$&0.341$^{\pm0.020}$&12.11$^{\pm0.27}$&341$^{\pm6.18}$&2.23$^{\pm0.12}$&26.3$^{\pm1.05}$&104/113  \\
24&X127& ---              &1.490$^{\pm0.032}$&103/106&---             &0.26$^{\pm0.001}$&1.2E-3$^{\pm1.0E-4}$&0.307$^{\pm0.023}$&12.18$^{\pm0.50}$&369$^{\pm6.20}$&2.13$^{\pm0.10}$&35.7$^{\pm1.29}$&114/99  \\
25&X128& ---              &1.769$^{\pm0.093}$& 94/106&---             &0.13$^{\pm0.002}$&1.2E-3$^{\pm1.1E-4}$&0.355$^{\pm0.020}$&12.25$^{\pm0.42}$&380$^{\pm7.58}$&2.29$^{\pm0.12}$&6.69$^{\pm0.71}$& 94/102  \\
27&X130& ---              &1.683$^{\pm0.033}$&127/101&---             &0.19$^{\pm0.004}$&1.0E-3$^{\pm1.1E-4}$&0.294$^{\pm0.020}$&12.41$^{\pm0.29}$&414$^{\pm8.23}$&2.62$^{\pm0.10}$&0.85$^{\pm0.01}$&115/90  \\
28&X131& ---              &1.631$^{\pm0.027}$&122/104&---             &0.21$^{\pm0.001}$&1.0E-3$^{\pm1.0E-4}$&0.278$^{\pm0.018}$&12.19$^{\pm0.22}$&428$^{\pm6.17}$&2.88$^{\pm0.11}$&22.3$^{\pm1.22}$&110/97  \\
29&X132& ---              &1.677$^{\pm0.039}$&104/96 &---             &0.20$^{\pm0.001}$&1.1E-3$^{\pm1.1E-4}$&0.245$^{\pm0.020}$&12.31$^{\pm0.32}$&441$^{\pm2.47}$&3.02$^{\pm0.10}$&7.58$^{\pm0.09}$& 91/97  \\
30&X133& ---              &1.605$^{\pm0.039}$&115/99 &---             &0.22$^{\pm0.001}$&1.0E-3$^{\pm1.0E-4}$&0.278$^{\pm0.030}$&12.07$^{\pm0.22}$&444$^{\pm4.19}$&3.19$^{\pm0.11}$&13.7$^{\pm0.39}$&108/97  \\
32&X135& ---              &1.353$^{\pm0.024}$&110/108&---             &0.22$^{\pm0.001}$&1.0E-3$^{\pm1.0E-4}$&0.273$^{\pm0.022}$&12.28$^{\pm0.52}$&449$^{\pm8.06}$&3.49$^{\pm0.12}$&10.1$^{\pm0.22}$&133/101  \\
33&X136& ---              &1.420$^{\pm0.026}$&102/105&---             &0.22$^{\pm0.001}$&1.0E-3$^{\pm1.0E-4}$&0.256$^{\pm0.021}$&12.24$^{\pm0.68}$&447$^{\pm8.21}$&3.50$^{\pm0.03}$&13.6$^{\pm0.02}$&130/104  \\
34&X137& ---              &1.473$^{\pm0.025}$&100/109&---             &0.20$^{\pm0.001}$&1.0E-3$^{\pm1.0E-4}$&0.259$^{\pm0.014}$&12.19$^{\pm0.50}$&448$^{\pm7.28}$&3.59$^{\pm0.10}$&10.8$^{\pm0.43}$&115/102  \\
35&X138& ---              &1.382$^{\pm0.037}$&104/105&---             &0.20$^{\pm0.001}$&1.0E-3$^{\pm1.0E-4}$&0.252$^{\pm0.020}$&12.31$^{\pm0.48}$&446$^{\pm6.17}$&3.66$^{\pm0.12}$&13.4$^{\pm0.32}$&103/98 \\
36&X139& ---              &1.360$^{\pm0.042}$&112/97 &---             &0.26$^{\pm0.001}$&1.0E-3$^{\pm0.9E-4}$&0.263$^{\pm0.044}$&12.17$^{\pm0.34}$&448$^{\pm5.12}$&12.2$^{\pm0.42}$&3.02$^{\pm0.12}$&110/89 \\
37&X140& ---              &1.354$^{\pm0.046}$&106/99 &---             &0.24$^{\pm0.001}$&1.0E-3$^{\pm0.9E-4}$&0.244$^{\pm0.014}$&12.16$^{\pm0.49}$&442$^{\pm4.78}$&3.72$^{\pm0.10}$&18.3$^{\pm0.45}$&105/90 \\
38&X141& ---              &1.414$^{\pm0.062}$& 79/88 &---             &0.20$^{\pm0.001}$&1.0E-3$^{\pm1.0E-4}$&0.292$^{\pm0.020}$&12.15$^{\pm0.42}$&442$^{\pm4.66}$&3.76$^{\pm0.14}$&4.80$^{\pm0.31}$& 78/82 \\
42&X145& ---              &1.396$^{\pm0.035}$& 88/101&---             &0.17$^{\pm0.001}$&1.0E-3$^{\pm0.8E-4}$&0.293$^{\pm0.023}$&12.28$^{\pm0.44}$&438$^{\pm4.33}$&3.80$^{\pm0.13}$&3.48$^{\pm0.28}$& 89/94 \\
43&X146& ---              &1.420$^{\pm0.050}$& 68/92 &---             &0.17$^{\pm0.001}$&1.0E-3$^{\pm1.0E-4}$&0.239$^{\pm0.020}$&12.14$^{\pm0.33}$&433$^{\pm4.48}$&3.94$^{\pm0.15}$&13.4$^{\pm0.34}$& 71/85 \\
44&X147& ---              &1.532$^{\pm0.068}$& 65/62 &---             &0.13$^{\pm0.001}$&3.9E-3$^{\pm1.1E-4}$&0.114$^{\pm0.030}$&11.88$^{\pm0.58}$&443$^{\pm5.15}$&3.97$^{\pm0.18}$&34.6$^{\pm1.03}$& 67/55 \\
45&X148& ---              &1.527$^{\pm0.067}$& 67/62 &---             &0.10$^{\pm0.001}$& 0.015$^{\pm 0.002}$&0.093$^{\pm0.009}$&12.52$^{\pm0.60}$&459$^{\pm6.93}$&3.75$^{\pm0.23}$&9.79$^{\pm0.32}$&115/85 \\
50&X153& ---              &1.411$^{\pm0.062}$& 95/98 &---             &0.09$^{\pm0.001}$&3.4E-3$^{\pm1.2E-4}$&0.257$^{\pm0.013}$&12.15$^{\pm0.50}$&354$^{\pm4.64}$&3.68$^{\pm0.24}$&3.52$^{\pm0.22}$&102/91 \\
54&X158&0.195$^{\pm0.011}$&2.072$^{\pm0.064}$&100/105&0.74$^{\pm0.03}$&1.04$^{\pm0.013}$& 0.011$^{\pm1.0E-4}$&0.233$^{\pm0.007}$&12.98$^{\pm0.60}$&321$^{\pm5.50}$&2.84$^{\pm0.23}$&8.95$^{\pm0.90}$&112/104 \\
56&X160&0.193$^{\pm0.019}$&2.016$^{\pm0.075}$&110/92 &0.46$^{\pm0.03}$&1.05$^{\pm0.018}$& 0.015$^{\pm1.3E-4}$&0.297$^{\pm0.030}$&12.22$^{\pm0.56}$&235$^{\pm5.18}$&1.84$^{\pm0.10}$&6.03$^{\pm0.23}$&111/90 \\
58&X162&0.221$^{\pm0.015}$&1.813$^{\pm0.080}$& 80/87 &0.54$^{\pm0.02}$&0.97$^{\pm0.017}$& 0.013$^{\pm 0.002}$&0.355$^{\pm0.029}$&12.26$^{\pm0.60}$&146$^{\pm4.56}$&1.56$^{\pm0.13}$&4.00$^{\pm0.20}$& 73/75 \\
62&X166&0.199$^{\pm0.022}$&1.945$^{\pm0.034}$&128/105&0.81$^{\pm0.02}$&1.08$^{\pm0.009}$& 0.031$^{\pm 0.003}$&0.293$^{\pm0.030}$&12.84$^{\pm0.69}$&134$^{\pm0.69}$&1.91$^{\pm0.12}$&1.64$^{\pm0.08}$&125/102 \\
63&X167&0.246$^{\pm0.009}$&1.603$^{\pm0.056}$&128/105&0.79$^{\pm0.02}$&1.03$^{\pm0.012}$& 0.037$^{\pm 0.013}$&0.369$^{\pm0.024}$&12.98$^{\pm0.68}$&128$^{\pm3.72}$&2.09$^{\pm0.11}$&0.78$^{\pm0.04}$&130/102 \\
65&X169&0.196$^{\pm0.009}$&1.933$^{\pm0.044}$&123/103&0.58$^{\pm0.02}$&0.91$^{\pm0.009}$& 0.010$^{\pm1.2E-4}$&0.271$^{\pm0.029}$&12.32$^{\pm0.48}$&132$^{\pm4.32}$&1.84$^{\pm0.10}$&4.26$^{\pm0.26}$&115/100 \\
67&X171&0.171$^{\pm0.010}$&2.025$^{\pm0.038}$&129/105&0.41$^{\pm0.02}$&0.88$^{\pm0.009}$& 0.026$^{\pm 0.002}$&0.263$^{\pm0.027}$&11.91$^{\pm0.45}$&129$^{\pm3.48}$&1.81$^{\pm0.09}$&3.43$^{\pm0.10}$&128/100 \\
69&X173&0.153$^{\pm0.011}$&2.014$^{\pm0.042}$& 91/102&0.38$^{\pm0.02}$&0.80$^{\pm0.009}$&9.4E-3$^{\pm1.1E-4}$&0.269$^{\pm0.026}$&12.47$^{\pm0.42}$&131$^{\pm4.60}$&1.87$^{\pm0.13}$&3.37$^{\pm0.09}$& 87/97 \\
72&X179&0.198$^{\pm0.023}$&1.986$^{\pm0.114}$& 85/73 &0.50$^{\pm0.04}$&0.76$^{\pm0.019}$& 4.7E-3$^{\pm 1.0E-4}$&0.259$^{\pm0.022}$&12.30$^{\pm0.53}$&112$^{\pm5.40}$&1.90$^{\pm0.25}$&5.10$^{\pm0.32}$& 84/70 \\
73&X185&0.156$^{\pm0.009}$&1.924$^{\pm0.033}$&128/105&0.36$^{\pm0.02}$&0.80$^{\pm0.008}$&4.0E-3$^{\pm0.9E-4}$&0.265$^{\pm0.032}$&12.26$^{\pm0.60}$& 79$^{\pm3.12}$&1.91$^{\pm0.24}$&3.10$^{\pm0.21}$&121/101 \\
74&X186&0.196$^{\pm0.010}$&1.851$^{\pm0.049}$& 87/101&0.54$^{\pm0.02}$&0.80$^{\pm0.010}$&8.4E-3$^{\pm1.2E-4}$&0.303$^{\pm0.042}$&12.48$^{\pm0.48}$& 95$^{\pm3.88}$&2.07$^{\pm0.14}$&0.32$^{\pm0.02}$& 96/98 \\
75&X187&0.174$^{\pm0.015}$&1.896$^{\pm0.059}$& 96/92 &0.34$^{\pm0.02}$&0.80$^{\pm0.013}$& 0.012$^{\pm 0.002}$&0.291$^{\pm0.023}$&12.60$^{\pm0.45}$&106$^{\pm4.44}$&2.04$^{\pm0.12}$&1.48$^{\pm0.03}$& 88/88 \\
76&X192&0.131$^{\pm0.021}$&1.708$^{\pm0.038}$&134/113&0.29$^{\pm0.01}$&1.02$^{\pm0.007}$& 0.016$^{\pm 0.003}$&0.258$^{\pm0.032}$&12.23$^{\pm0.60}$&129$^{\pm4.04}$&1.98$^{\pm0.13}$&3.74$^{\pm0.33}$&136/110 \\
77&X193&0.187$^{\pm0.007}$&1.946$^{\pm0.051}$& 79/104&0.89$^{\pm0.02}$&1.20$^{\pm0.011}$& 0.016$^{\pm 0.004}$&0.267$^{\pm0.039}$&12.35$^{\pm0.55}$&131$^{\pm4.83}$&1.83$^{\pm0.10}$&3.94$^{\pm0.29}$& 72/101 \\
78&X195&0.252$^{\pm0.028}$&1.963$^{\pm0.052}$&122/103&0.42$^{\pm0.01}$&1.18$^{\pm0.011}$& 0.016$^{\pm 0.003}$&0.292$^{\pm0.029}$&12.50$^{\pm0.60}$&125$^{\pm4.09}$&1.19$^{\pm0.15}$&5.96$^{\pm0.24}$&124/100 \\
79&X196&0.174$^{\pm0.012}$&1.827$^{\pm0.073}$& 88/79 &0.83$^{\pm0.05}$&1.31$^{\pm0.024}$& 0.019$^{\pm 0.003}$&0.295$^{\pm0.043}$&12.17$^{\pm0.48}$&125$^{\pm4.92}$&2.13$^{\pm0.14}$&2.26$^{\pm0.18}$& 97/88 \\
83&X209&0.172$^{\pm0.013}$&2.123$^{\pm0.056}$&114/96 &0.40$^{\pm0.02}$&0.77$^{\pm0.011}$& 0.015$^{\pm 0.002}$&0.285$^{\pm0.033}$&12.22$^{\pm0.50}$& 94$^{\pm4.53}$&1.16$^{\pm0.12}$&2.18$^{\pm0.04}$&114/92 \\
84&X210&0.164$^{\pm0.013}$&2.087$^{\pm0.062}$& 75/79 &0.54$^{\pm0.03}$&0.78$^{\pm0.012}$& 0.012$^{\pm 0.001}$&0.208$^{\pm0.029}$&12.85$^{\pm0.68}$& 94$^{\pm3.86}$&1.86$^{\pm0.44}$&2.71$^{\pm0.05}$& 76/76 \\
85&X213&0.168$^{\pm0.018}$&2.066$^{\pm0.085}$& 91/88 &0.35$^{\pm0.03}$&0.64$^{\pm0.014}$& 0.014$^{\pm 0.003}$&0.244$^{\pm0.020}$&12.18$^{\pm0.60}$& 94$^{\pm4.43}$&1.58$^{\pm0.12}$&2.06$^{\pm0.05}$& 92/83 \\
86&X215&0.188$^{\pm0.006}$&1.866$^{\pm0.031}$&124/108&0.62$^{\pm0.01}$&0.79$^{\pm0.006}$&8.9E-3$^{\pm1.1E-4}$&0.202$^{\pm0.019}$&12.13$^{\pm0.51}$& 50$^{\pm2.40}$&1.93$^{\pm0.28}$&1.83$^{\pm0.07}$&124/105 \\
89&X219&0.208$^{\pm0.020}$&2.133$^{\pm0.095}$& 83/89 &0.45$^{\pm0.03}$&0.75$^{\pm0.015}$&7.8E-3$^{\pm1.0E-4}$&0.167$^{\pm0.014}$&12.19$^{\pm0.59}$& 90$^{\pm3.42}$&1.75$^{\pm0.27}$&6.43$^{\pm0.15}$& 87/86 \\
93&X227&0.196$^{\pm0.032}$&1.698$^{\pm0.133}$& 69/83 &0.30$^{\pm0.03}$&0.58$^{\pm0.019}$&3.8E-3$^{\pm1.0E-4}$&0.252$^{\pm0.020}$&12.27$^{\pm0.49}$& 48$^{\pm2.50}$&2.24$^{\pm0.22}$&1.21$^{\pm0.22}$& 61/80 \\
95&X229&0.174$^{\pm0.010}$&1.964$^{\pm0.044}$& 85/105&0.28$^{\pm0.01}$&0.56$^{\pm0.006}$&5.6E-3$^{\pm1.1E-4}$&0.186$^{\pm0.022}$&12.42$^{\pm0.52}$& 52$^{\pm2.12}$&1.86$^{\pm0.29}$&2.28$^{\pm0.19}$&101/104 \\
96&X230&0.169$^{\pm0.009}$&1.878$^{\pm0.048}$&123/103&0.41$^{\pm0.02}$&0.63$^{\pm0.008}$&9.7E-3$^{\pm0.9E-4}$&0.197$^{\pm0.026}$&12.37$^{\pm0.44}$& 49$^{\pm2.13}$&2.15$^{\pm0.14}$&0.16$^{\pm0.04}$&119/101 \\
97&X231&0.196$^{\pm0.010}$&1.848$^{\pm0.055}$&120/101&0.39$^{\pm0.01}$&0.53$^{\pm0.007}$&7.1E-3$^{\pm1.0E-4}$&0.200$^{\pm0.020}$&12.20$^{\pm0.65}$& 68$^{\pm2.33}$&2.39$^{\pm0.29}$&1.68$^{\pm0.20}$&116/98 \\
\hline
\end{tabular}
%\noindent{
\leftline{X (=6203690) is the initial part of Observation IDs. A fixed N$_{\rm H}$=$9\times10^{21}~cm^{-2}$ is used for both set of models. }
\leftline{The units of T$_{\rm in}$ in keV; F$_{\rm DBB}$ and F$_{\rm PL}$ fluxes in $10^{-10}~{\rm erg}~{\rm cm}^{-2}~{\rm s}^{-1}$, $\dot{m}_d$ and $\dot{m}_h$ in Eddington rate ($\dot{M}_{\rm Edd}$); M$_{\rm BH}$ in M$_\odot$; X$_s$ in Schwarzschild radius. }
%\leftline{$\nu_{QPO}$ in Hz; $N_{\rm H}$ in $10^{22}~{\rm cm}^{-2}$; $T_{in}$ in keV; model fluxes diskbb ($F_{\rm DBB}$) and pexrav ($F_{\rm pexr}$) in $10^{-8}~{\rm erg}~{\rm cm}^{-2}~{\rm s}^{-1}$.}
\end{table*}

\end{document}